\documentclass[pra,aps,twocolumn,showpacs,superscriptaddress,nofootinbib]{revtex4-1}
\usepackage{graphicx} \usepackage{dcolumn} 
\usepackage{bm}
\usepackage[bbgreekl]{mathbbol}
\usepackage{amsmath}
\usepackage{xr}
\usepackage{graphicx}
\usepackage{color}

\def\s{\hat{\sigma}}

\begin{document}

\title{Driven-dissipative phase separation in free-space atomic ensembles}
\author{D. Goncalves}
\address{ICFO-Institut de Ciencies Fotoniques, The Barcelona Institute of Science and Technology, 08860 Castelldefels (Barcelona), Spain.}
\author{L. Bombieri}
\address{ICFO-Institut de Ciencies Fotoniques, The Barcelona Institute of Science and Technology, 08860 Castelldefels (Barcelona), Spain.}
\affiliation{Institute for Theoretical Physics, University of Innsbruck, 6020 Innsbruck, Austria}
\affiliation{Institute for Quantum Optics and Quantum Information of the Austrian Academy of Sciences, 6020 Innsbruck, Austria}
\author{G. Ferioli}
\address{Universite Paris-Saclay, Institut d’Optique Graduate School, CNRS, Laboratoire Charles Fabry, 91127, Palaiseau, France.}
\author{S. Pancaldi}
\address{Universite Paris-Saclay, Institut d’Optique Graduate School, CNRS, Laboratoire Charles Fabry, 91127, Palaiseau, France.}
\author{I.~Ferrier-Barbut}
\address{Universite Paris-Saclay, Institut d’Optique Graduate School, CNRS, Laboratoire Charles Fabry, 91127, Palaiseau, France.}
\author{A. Browaeys}
\address{Universite Paris-Saclay, Institut d’Optique Graduate School, CNRS, Laboratoire Charles Fabry, 91127, Palaiseau, France.}
\author{E. Shahmoon}
\address{Department of Chemical \& Biological Physics, Weizmann Institute of Science, Rehovot 7610001, Israel.}
\author{D. E. Chang}
\address{ICFO-Institut de Ciencies Fotoniques, The Barcelona Institute of Science and Technology, 08860 Castelldefels (Barcelona), Spain.}
\address{ICREA-Instituci{\'o} Catalana de Recerca i Estudis Avan{\c c}ats, 08015 Barcelona, Spain.}

\date{\today}
\begin{abstract}
The driven Dicke model, wherein an ensemble of atoms is driven by an external field and undergoes collective spontaneous emission due to coupling to a leaky cavity mode, is a paradigmatic example of a system exhibiting a driven-dissipative phase transition as a function of driving strength. Recently, a similar phenomenon was experimentally observed, not in a cavity setting, but rather in a free-space atomic ensemble. The reason why similar behavior should emerge in free space is not obvious, as the system interacts with a continuum of optical modes, which encodes light propagation effects. Here, we present and solve a simple model to explain the behavior of the free-space system, based on the one-dimensional Maxwell-Bloch equations. On one hand, we show that a free-space ensemble at a low optical depth can exhibit similar behavior as the cavity system, as spatial propagation effects are negligible. On the other hand, in the thermodynamic limit of large atom number, we show that certain observables such as the transmittance or the atomic excited population exhibit non-analytic behavior as a function of the driving intensity, reminiscent of a phase transition. However, a closer analysis reveals that the atomic properties are highly inhomogeneous in space, and based on this we argue that the free-space system does not undergo a phase transition but rather a ``phase separation,'' roughly speaking, between saturated and unsaturated regions. 
 
\end{abstract}

\maketitle

\section{Introduction}

\begin{figure}[b!]
    \centering
    \includegraphics[width=0.96\linewidth]{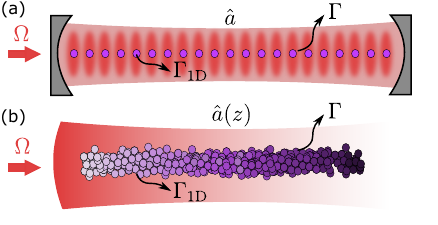}
    \caption{(a) Sketch of the driven-Dicke model, where an ensemble of $N$ two-level atoms couples identically to a single cavity mode $\hat{a}$. The ensemble is driven by a coherent field with strength $\Omega$ and decays collectively into the lossy cavity at a rate $\Gamma_{\rm 1D}$. In realistic setups, atoms can also decay independently outside of the cavity, i.e. into the $4\pi$ continuum of modes, at a rate $\Gamma$. Here, the system is permutational invariant and spatially homogeneous. (b) Sketch of the Maxwell-Bloch model representing an ensemble in free space, where atoms couple to the driving mode at rate $\Gamma_{\rm 1D}$ and to the continuum of modes at a rate $\Gamma$. Here, light propagation effects prevent the field $\hat{a}(z)$ and atomic operators $\s^\alpha(z)$ from being spatially uniform in general. }
    \label{fig:Sketch_Fig1}
\end{figure}

The driven-dissipative Dicke model stands as a paradigmatic example of a driven, open quantum system. It represents a collection of coherently driven, two-level systems uniformly coupled to a single, lossy electromagnetic field mode. Despite its apparent simplicity, it approximately describes an atomic ensemble coupled to a cavity \cite{HeppLieb1973, Argawall1977, Ritsch2013, Norcia2016}, like in Fig.~\ref{fig:Sketch_Fig1}a, and displays several rich features that have stimulated sustained research. A notable aspect of this model is the existence of a driven-dissipative phase transition~\cite{Carmichael_1977, Carmichael1980, Schneider2002, Tudela2013, Kirton2017, Hannukainen2018, Kirton2019, Barberena_2019, Somech2022}, governed by the parameter $\beta = 2\Omega/N\Gamma_{\rm 1D}$, with $\Omega$ and $\Gamma_{\rm 1D}$ being the coherent driving strength and atomic emission rate into the cavity, respectively. The system exhibits a magnetized ($\beta<1$) and non-magnetized ($\beta>1$) phase, with the latter manifesting non-ergodic dynamics \cite{Munoz2019, Barberena_2019} and so-called time crystalline behavior \cite{Drumond1978,Iemini2018, Zhu_2019}. The magnetized phase exhibits spin squeezing \cite{Puri1979, Tudela2013}, which is known to be a valuable resource in quantum metrology. In realistic cavity setups, atoms can also decay independently into free space at a rate $\Gamma$, leading to first-order phase transition with modified properties, like a shifted critical value of $\beta$ \cite{Drumond1978,Carmichael_1977, Carmichael1980, Walls78, Gegg_2018}.\\

Surprisingly, a recent experiment \cite{Ferioli2023} has shown that an elongated ensemble of atoms in \textit{free space}, driven along its main axis, exhibits features that look remarkably similar to the driven-dissipative Dicke model, even though there are no cavity mirrors. For example, the magnetization was observed to be suppressed as a function of driving strength, in a way resembling the phase transition in the Dicke model. The atomic emission also appeared to be superradiant, scaling quadratically with system size for strong driving. However, formally, the cavity and free space systems appear quite different. The Dicke model involves a discrete cavity mode and thus has no spatial field degrees of freedom, resulting in permutational invariance of atomic observables and conservation of total angular momentum.~Contrarily, a free space atomic ensemble couples to a continuum of field modes, and indeed spatiotemporal propagation effects are behind various well-known phenomena in such systems \cite{ Phillips2001, Liu2001, Fleischhauer2005, Firstenberg2013}. \\

Here, we examine the behavior of a driven elongated atomic ensemble in free space to elucidate the possible existence of a phase transition, and its connection to that of the iconic ``zero-dimensional" driven Dicke model. To this end, we present the model illustrated in Fig.~\ref{fig:Sketch_Fig1}b, which is based on the well-established 1D Maxwell-Bloch (M-B) equations for light propagation in dilute atomic clouds \cite{McCall1967, Arecci1965, Haake1979, Podler1979, gross1982, Molmer1995, Hammerer2010, Gorshkov2011}. First, taking a discrete version of this theory, we show the atomic master equation is identical to the cavity problem in terms of driving and dissipation, but features an extra coherent (dipole-dipole) interaction encoding light propagation. This term captures the key difference between cavity and free-space scenarios and suggests their behavior converges when the atomic system is approximately spatially uniform~(e.g., for low optical depths or strong saturation). Second, solving the M-B model in the thermodynamic limit within the mean-field approximation, we find a non-analytic behavior in the light transmission and magnetization, and identify the parameter governing an apparent ``transition''. A closer inspection, however, reveals that the system exhibits what we term a ``phase separation'', where different spatial regions of the ensemble acquire different properties~(saturated vs. magnetized) due to the spatial degree of freedom. Finally, we compare our M-B model with the experimental results of Ref.~\cite{Ferioli2023}, obtaining a good agreement. 

We highlight the complementary work from Ref.~\cite{Argawal_2024}, which directly models a three-dimensional atomic ensemble, and arrives at similar conclusions to ours. Together, these works offer significant insight into the physics of driven-dissipative free-space atomic ensembles and the manifestation of cavity-like behavior.

\section{Driven-Dissipative Dicke model}

To facilitate the comparison with the free space results discussed later, we briefly recall the driven-dissipative Dicke model of $N$ two-level atoms interacting identically with a single-mode cavity, and summarize its main features. Assuming that the cavity mode is lossy, and starting from the full atom-light master equation, one can trace out the cavity field within a Born-Markov approximation to obtain a master equation for the atomic density matrix $\hat{\rho}$ that reads \cite{Argawall1977, Drumond1978, Carmichael_1977, Carmichael1980, Schneider2002}
\begin{equation}
    \dot{\hat{\rho}} = -\frac{i}{\hbar}\left[\hat{H}_{\rm drive},\hat{\rho}\right]+ \mathcal{L}_{\Downarrow}[\hat{\rho}].
    \label{eq:master_equation_Driven_Dicke_model}
\end{equation}
The Hamiltonian term 
\begin{equation}
    \hat{H}_{\rm drive} = \frac{\hbar\Omega}{2} (\hat{S}_++\hat{S}_-),
\end{equation}
describes a resonant coherent field driving the ensemble with Rabi amplitude $\Omega$. Here, $\hat{S}_\pm=\sum_{n=1}^N\hat{\sigma}_n^\pm$ are collective spin raising and lowering operators, which can be written in terms of the raising/lowering operators of the individual two-level atoms $\s_n^{\pm}$. 
The Liouvillian
\begin{equation}
    \mathcal{L}_{\Downarrow}[\rho]=\frac{\Gamma_{\rm 1D}}{2}\left( 2\hat{S}_-\hat{\rho} \hat{S}_+- \{\hat{\rho}, \hat{S}_+\hat{S}_-\}  \right)
    \label{eq:Lindbladian_Dicke}
\end{equation}
captures the collective atomic emission at a rate $\Gamma_{\rm 1D}$, which stems from the coupling to the lossy cavity.

The master equation (\ref{eq:master_equation_Driven_Dicke_model}) is invariant under the permutation of two atoms and conserves total angular momentum $\langle \hat{S}^2\rangle$. These symmetries allow $\hat{\rho}$ to be expressed in terms of a limited number of collective angular momentum states $|j,m\rangle$, which satisfy
\begin{subequations}
\begin{equation}
        \hat{S}^2|j,m\rangle =j(j+1)\ |j,m\rangle,
            \vspace{-1em}
\end{equation}
\begin{equation}
        \hat{S}_z|j,m\rangle = m\ |j,m\rangle,
\end{equation}
\end{subequations}
with $\hat{S}_\alpha =\sum_{n=1}^N\s_n^\alpha /2$, $j\in[0,N/2]$ and $m\in[-j,j]$. The conservation of angular momentum breaks the Hilbert space into different $j$ manifolds, which makes dynamics efficient to solve.


 \begin{figure*}[ht]
    \centering
    \includegraphics[width=0.99\textwidth]{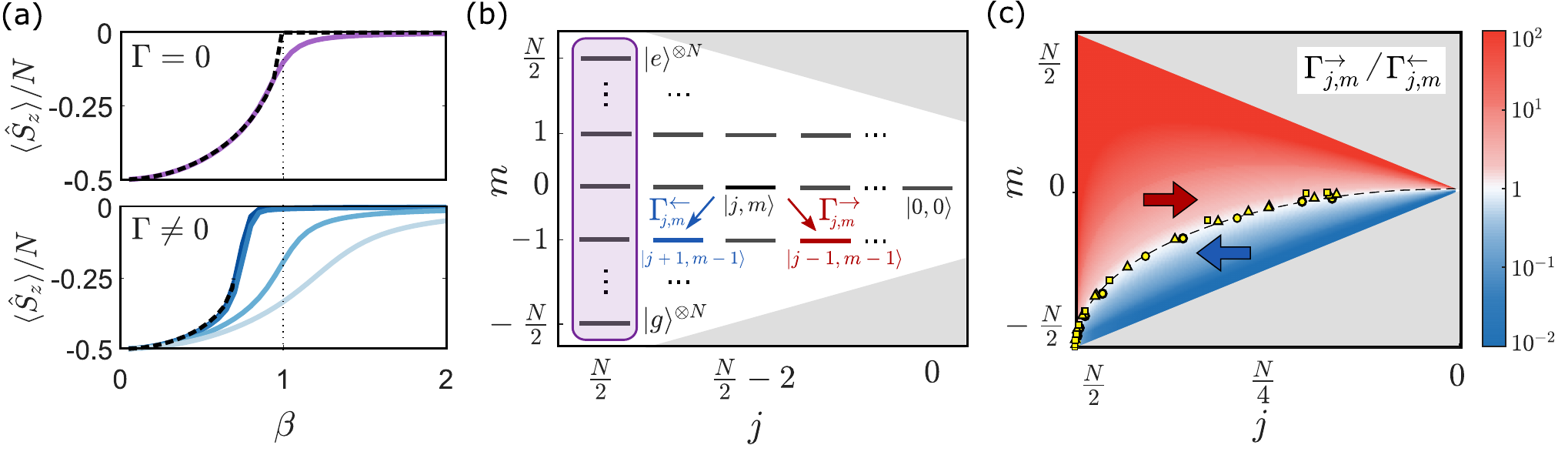} 
    \vspace{-0.5em}
    \caption{(a) Normalized steady-state magnetization $\langle \hat{S}_z\rangle/N$ as a function of $\beta=2\Omega/N\Gamma_{\rm 1D}$ for $N=30$ atoms. Top: Solid line is a numerical simulation of the Dicke model [Eq.~(\ref{eq:master_equation_Driven_Dicke_model})], while the dashed line is the mean-field prediction of Eq.~(\ref{eq:Mean_field_magnetization_Dicke}). Bottom: Solid lines are a simulation of the Cavity model [Eq.~(\ref{eq:Cavity_model})] for values of local dissipation strengths of $\Gamma=[0.5, 1, 5, 10]\Gamma_{\rm 1D}$ (dark to lighter blue), while the dashed line is Eq.~(\ref{eq:self_consistent_mean_field_solutions}). (b) Schematic ordering of the $|j,m\rangle$ states by their quantum numbers $m$ and $j$. Without local dissipation, $j$ is conserved and any dynamics starting from the ground state $|g\rangle^{\otimes N}=|j=N/2,m=-N/2\rangle$ is confined to the purple box. Instead, local dissipation allows for transitions $\Gamma^{\rightarrow}_{j,m}$ and $\Gamma^{\leftarrow}_{j,m}$ that modify $j$. (c) Ratio between the angular momentum-changing decay rates $\Gamma^{\rightarrow}_{j,m}/\Gamma^{\leftarrow}_{j,m}$ for different $j$ and $m$, which balance at the dashed line given by Eq.~(\ref{eq:relation_from_lateral_balance}). The yellow points are the results of numerical simulations of Eqs.~(\ref{eq:Cavity_model}) for different driving strengths, where we convert the expectation values of angular momentum operators to $j,m$ using the formulas $m\approx\langle \hat{S}_z\rangle$ and $j(j+1)\approx\langle \hat{S}^2\rangle=\langle \hat{S}_{x}^2+\hat{S}_{y}^2+\hat{S}_{z}^2\rangle$. Here, $N=30$ and $\Gamma=[0.1, 1, 10]\Gamma_{\rm 1D}$ (squares, points and triangles).}
\label{fig:Magnetization_Dicke_cavity}
\end{figure*}

\subsection{Driven-dissipative phase transition}
As discussed in the introduction, the competition between the coherent driving $\Omega$ and the collective dissipation $\Gamma_{\rm 1D}$ in Eq.~(\ref{eq:master_equation_Driven_Dicke_model}) yields a driven-dissipative phase transition \cite{Drumond1978, Carmichael_1977, Carmichael1980, Schneider2002, Tudela2013, Kirton2017, Hannukainen2018, Kirton2019, Barberena_2019, Somech2022}.  As a function of the ratio $\beta = 2\Omega/N\Gamma_{\rm 1D}$, the steady-state solution of the density matrix features a magnetization that goes from non-zero ($\beta<1$) to zero ($\beta>1$), a transition that can already be seen at the mean-field level. In the following, we focus on the $j=N/2$ manifold, which contains the state 
$|j=N/2,m=-N/2\rangle$ with all atoms unexcited~(the typical initial state). Keeping only the terms of order $\mathcal{O}(N^2)$, the mean-field Heisenberg equations for the collective spin operators in the steady-state read
\begin{subequations}\label{eq:Mean_field_Heisenberg_Dicke}
    \begin{equation}
        \Omega \langle \hat{S}_z\rangle  = \Gamma_{\rm 1D}\langle \hat{S}_z\rangle \langle \hat{S}_y\rangle,
    \vspace{-1em}
    \end{equation}
    \begin{equation}
        \Omega \langle \hat{S}_y\rangle = \Gamma_{\rm 1D}(\langle \hat{S}^2\rangle -\langle \hat{S}_z \rangle^2),
    \end{equation}
\end{subequations}
where $\langle \hat{S}_x\rangle=0$. To solve Eqs.~(\ref{eq:Mean_field_Heisenberg_Dicke}), one usually exploits the conservation of angular momentum to substitute $\langle \hat{S}^2\rangle= N(N/2+1)/2\approx N^2/4$, yielding 
\begin{equation}
     \langle \hat{S}_z\rangle  = -\frac{N}{2}\sqrt{1-\beta^2},\qquad \beta<1.
    \label{eq:Mean_field_magnetization_Dicke}
\end{equation}
Figure \ref{fig:Magnetization_Dicke_cavity}a shows the magnetization of the system $\langle\hat{S}_z\rangle$ as a function of $\beta$. The mean-field prediction in Eq.~(\ref{eq:Mean_field_magnetization_Dicke}) is consistent with a full master equation simulation of the Dicke model (\ref{eq:master_equation_Driven_Dicke_model}). The non-analytic behavior of $\langle\hat{S}_z\rangle$ at the critical point $\beta_c=1$ is a hallmark of the phase transition. We note this mean-field result is supported by other methods, such as numerics \cite{Hannukainen2018, Barberena_2019, Tudela2013}, exact solution \cite{Drumond1978, Puri1979}, or an analysis based on approximate eigenstates \cite{Somech2022}. Additional hallmarks of the phase transition are the steady-state density matrix going from a pure to completely mixed state~\cite{Tudela2013, Puri1979} and from squeezed to unsqueezed~\cite{Tudela2013, Dalla_Torre_2013} at the same critical point.

The unmagnetized phase $(\beta>1)$ also features superradiant atomic emission, with an emission rate scaling quadratically with atom number. Indeed, the atomic steady-state for $\beta>1$ is completely mixed within the $j=N/2$ manifold, which implies $\Gamma_{\rm 1D}\langle \hat{S}^+\hat{S}^-\rangle=\Gamma_{\rm 1D}\text{Tr}(\hat{S}^+\hat{S}^-)/(N+1)\propto N^2$. Notably, this superradiance is preserved at arbitrarily large values of $\beta$ thanks to angular momentum conservation.

Although not central to the discussion, we point out that even if the steady-state solution for $\beta>1$ is a mixed density matrix, a large system can undergo non-ergodic dynamics with persistent oscillations \cite{Munoz2019}, an effect which has been termed as a boundary time crystal \cite{Iemini2018,Drumond1978,Zhu_2019}.

\subsection{Realistic Cavity model}\label{Sec:Cavity_model}

In realistic cavity QED setups, the cavity is not closed and atoms can also emit into the continuum of $4\pi$ modes. The typical assumption in quantum optics consists in treating these other modes as independent emission, which modifies the original Dicke model by a local dissipation term $\mathcal{L}_{\downarrow}[\hat{\rho}]$, such that
\begin{subequations}\label{eq:Cavity_model}
\begin{equation}
        \dot{\hat{\rho}} = -\frac{i}{\hbar}\left[\hat{H}_{\rm drive},\hat{\rho}\right]+ \mathcal{L}_{\Downarrow}[\hat{\rho}]+ \mathcal{L}_{\downarrow}[\hat{\rho}],
            \vspace{-1em}
\end{equation}
    \begin{equation}
   \mathcal{L}_{\downarrow}[\hat{\rho}]=\frac{\Gamma}{2}\sum_{n=1}^N\left( 2\s^-_n\hat{\rho} \s^+_n- \{\hat{\rho}, \s^+_n\s^-_n\}  \right).
    \label{eq:Lindbladian_cavity}
\end{equation}
\end{subequations}
Equations (\ref{eq:Cavity_model}) will be referred to as the Cavity model, where atoms decay collectively into the cavity mode at a rate $\Gamma_{\rm 1D}$, and independently into free space at a rate $\Gamma$. 

The local dissipation term $\mathcal{L}_{\downarrow}[\hat{\rho}]$ breaks the conservation of angular momentum and allows the system to explore multiple $j$ manifolds during its evolution. Even so, Eqs.~(\ref{eq:Cavity_model}) remain invariant under the exchange of two atoms. This permutational symmetry can be exploited to aid numerical simulations, as it significantly reduces the Hilbert space size, only requiring $\mathcal{O}(N^3)$ elements instead of $\mathcal{O}(2^N)$ \cite{Sarkar_1987, Shammah2018, Minghui,Geremiah}.

\subsection{Dynamics in the Cavity model}\label{Sec:lateral_decay_rates}

The non-conservation of angular momentum makes the system depart from the initial $j=N/2$ manifold. To understand why, it is convenient to rewrite the cavity master Eq.~(\ref{eq:Cavity_model}) in terms of the angular momentum states by projecting the density matrix $\hat{\rho}$ into the $|j,m\rangle$ basis, 
\begin{equation}
    \langle j',m-1|\dot{\hat{\rho}}| j',m-1\rangle=\sum_{j,m}\Gamma^{\alpha}_{j,m}\langle j,m|\hat{\rho}|j,m\rangle+...
\end{equation}
This reveals the rates $\Gamma_{j,m}^{\alpha}$ at which population is transferred from the state $|j,m\rangle$ into the states $|j',m-1\rangle$ via spontaneous emission, where $\alpha=\{\leftarrow,\rightarrow,\downarrow\}$ indicates whether the angular momentum $j'$ of the final state has increased or decreased by one unit or stayed the same, as illustrated in Fig.~\ref{fig:Magnetization_Dicke_cavity}b. In the following, we focus exclusively on the angular momentum-changing decay rates $\Gamma^{\leftarrow}_{j,m}$ and $\Gamma^{\rightarrow}_{j,m}$, which change the value of $j$ by $+1$ and $-1$, respectively,  and whose explicit form can be found in the literature to be \cite{Shammah2018}
\begin{subequations}\label{eq:rate_in}
          \begin{equation}\label{eq:Decay_right}
    \Gamma^{\leftarrow}_{j,m}=\Gamma\frac{(j-m+1)(j-m+2)(\frac{N}{2}-j)}{2(j+1)(2j+1)},
    \vspace{-0.5em}
    \end{equation} 
    \begin{equation}\label{eq:Decay_left}
    \Gamma^{\rightarrow}_{j,m}=\Gamma\frac{(j+m-1)(j+m)(j+1+\frac{N}{2})}{2j(2j+1)}.
    \end{equation}
\end{subequations}
Figure \ref{fig:Magnetization_Dicke_cavity}c illustrates how the ratio between these decay rates depends on $j$ and $m$. 
In the red region, $\Gamma^{\rightarrow}_{j,m}\gg\Gamma^{\leftarrow}_{j,m}$ and a given state evolves toward lower $j$ manifolds. The opposite happens in the blue region, where the angular momentum increases. \\

Intuitively, in the steady state, one expects the system to lie on the white curve where these decay rates balance. Let us treat $j$ and $m$ as continuous variables and consider the limit $N\gg1$. Imposing $\Gamma^{\rightarrow}_{j,m}=\Gamma^{\leftarrow}_{j,m}$, one finds a universal relation between angular momentum and magnetization in the steady state
\begin{equation}
    j =\frac{N}{2}\sqrt{1-(1+2m/N)^2},
\label{eq:relation_from_lateral_balance}
\end{equation}
which interestingly is independent of $\Gamma$ (provided it is not exactly zero). We supplement this intuitive argument with full master equation simulations. In particular, in Fig.~\ref{fig:Magnetization_Dicke_cavity}c, we plot with points the calculated steady-state values of $\langle \hat{S}_z\rangle$ and $\langle \hat{S}^2\rangle$ for $N=30$ atoms and various ratios of $\Gamma/\Gamma_{\rm 1D}$. We find that, after associating the expectation values of the observables with $m\approx \langle \hat{S}_z\rangle$ and $j(j+1)\approx\langle \hat{S}^2\rangle$, the steady-state values indeed lie on the predicted universal curve of Eq.~(\ref{eq:relation_from_lateral_balance}).\\

As an aside, one can use similar arguments to derive conditions under which the steady state in the Cavity model becomes mixed. Indeed, since Eq.~(\ref{eq:Cavity_model}) never creates coherences between different $j$ manifolds, one expects a pure state only if the population never leaves the initial $j$ manifold, or equivalently, if $\Gamma^{\rightarrow}_{j,m}<\Gamma^{\leftarrow}_{j,m}$. Let us focus on states close to the $j=N/2$ manifold, such that $j=\frac{N}{2}-\epsilon_j$ with $\epsilon_j\sim \mathcal{O}(1)$. Using again Eqs.~(\ref{eq:rate_in}) in the $N\gg 1$ limit, this inequality relation between the decay rates yields
\begin{equation}
    n\lesssim \frac{1}{\sqrt{N}},
    \label{eq:condition_spread}
\end{equation}
where $n=m/N+0.5$ is the fraction of excited population. Interestingly, the condition (\ref{eq:condition_spread}) cannot be satisfied in the thermodynamic limit $(N\rightarrow \infty)$ for a non-zero $n$. This is in stark contrast with the Dicke model, where the system is always pure in the magnetized phase $(\beta<1)$.

\subsection{Phase transition in the Cavity model}
A phase transition still persists in the presence of local dissipation, albeit with modified properties. Let us consider a small amount of local dissipation $\Gamma<N\Gamma_{\rm 1D}$ in Eq.~(\ref{eq:Cavity_model}). Since independent emission only introduces corrections of order $\mathcal{O}(N)$ in the Heisenberg equations, Eqs.~(\ref{eq:Mean_field_Heisenberg_Dicke}) remain valid. However, now $\langle \hat{S}^2\rangle\not=N^2/4$ because angular momentum is not conserved. Instead, we substitute the value of angular momentum using Eq.~(\ref{eq:relation_from_lateral_balance}), where we again approximate the values of $j$ and $m$ by the expectation values of $\langle \hat{S}^2\rangle$ and $\langle \hat{S}_z\rangle$, respectively. This self-consistent mean-field theory yields three solutions, from which only $\langle \hat{S}^z\rangle=0$ and 
    \begin{equation}
         \langle \hat{S}_z\rangle =-\frac{N}{4}-\frac{N}{4}\sqrt{1-2\beta^2},\qquad \beta<\frac{1}{\sqrt{2}},
        \label{eq:self_consistent_mean_field_solutions}
    \end{equation}
are stable. Equation (\ref{eq:self_consistent_mean_field_solutions}) indicates a first-order phase transition at a shifted critical point $\beta_c'=1/\sqrt{2}$ compared to the original Dicke model ($\beta_c=1$). Figure \ref{fig:Magnetization_Dicke_cavity}a shows how Eq.~(\ref{eq:self_consistent_mean_field_solutions}) agrees with a master equation simulation of the Cavity model (\ref{eq:Cavity_model}) for a fixed $N$ and sufficiently small $\Gamma$. On the other hand, for increasing $\Gamma$~(but fixed $N$), the transition from magnetized to unmagnetized is displaced to larger values of $\beta$, far away from the $\beta'_c$ prediction in Eq.~(\ref{eq:self_consistent_mean_field_solutions}). Again, these findings are supported by other methods such as a factorization ansatz \cite{Walls78, Carmichael1980, Drumond1978} and numerics \cite{Sarkar_1987, Gegg_2018}.

\section{Ensemble in free space}

We now introduce our minimal model for the experiment reported in Ref.~\cite{Ferioli2023}, where an elongated ensemble in free space is driven quasi-uniformly by a focused laser beam~(approximately a Gaussian transverse mode). Our goal is to describe the dynamics of both the atomic operators and the light propagating in this quasi-one-dimensional mode. To this end, we utilize the 1D Maxwell-Bloch~(M-B) equations, which constitute the predominant theory to treat these systems \cite{McCall1967, Arecci1965, Haake1979, Podler1979, gross1982, Molmer1995, Hammerer2010, Gorshkov2011}. We first introduce these equations in a somewhat different form to what appears in textbooks -- in particular taking the atoms as discrete point-like scatterers -- before showing their equivalence to the standard form of the M-B equations, where atoms appear as a continuous density. This discrete formulation has the advantage that it highlights the mathematical difference with the Cavity model, once the field is integrated out.

\subsection{Maxwell-Bloch model}

We start from the atom-light Hamiltonian of $N$ atoms coupled to a one-dimensional continuum of propagating photons in free space, described by the field operator $\hat{a}(z)$. This field represents the right-propagating mode by which atoms are driven and probed. The system Hamiltonian, including both atoms and photons, reads \cite{Giuseppe2018,Fan2009,Fan2005,Giuseppe2022, Calajo2020}
\begin{multline}
        \hat{H}=-i\hbar c\int \text{d}z\left[\hat{a}^\dagger(z)\partial_z\hat{a}(z) \right]+\\        
        +\hbar\sqrt{\frac{c\Gamma_{\rm 1D}}{2}}\sum_{n=1}^N\left[\hat{a}^\dagger(z_n)\s^-_n+h.c.\right].
        \label{eq:Full_wQED_Hamiltonian}
\end{multline}
The first term describes the energy of the free-propagating fields, where $\hat{a}^\dagger(z)$ and $\hat{a}(z)$ create and annihilate a right-propagating photon at position $z$. They satisfy standard bosonic commutation relations $[\hat{a}(z),\hat{a}^\dagger(z')]=\delta(z-z')$. The second term in Eq.~(\ref{eq:Full_wQED_Hamiltonian}) accounts for the interaction between atoms at discrete positions $z_n$ and photons, where we have assumed an identical, frequency-independent coupling rate $\Gamma_{\rm 1D}$. 
 
Next, we use Eq.~(\ref{eq:Full_wQED_Hamiltonian}) to calculate the Heisenberg equations of motion for the field and atomic operators
\begin{subequations}\label{eq:Heisenberg_wQED_equations}
\begin{equation}
    \left[\frac{\partial}{\partial z}+\frac{1}{c}\frac{\partial}{\partial t}\right]\hat{a}(z)=-i\sqrt{\frac{\Gamma_{\rm 1D}}{2c}}\sum_{n=1}^{N}\s^-_n\delta(z-z_n),
    \label{eq:Heisenberg_wQED_fields}
\end{equation}
    \begin{equation}
    \frac{\partial}{\partial t}\s^-_n=i\sqrt{\frac{c\Gamma_{\rm 1D}}{2}}\hat{a}(z_n)\s_n^z,
\end{equation}
    \begin{equation}
     \frac{\partial}{\partial t}\s^z_n=i\sqrt{2c\Gamma_{\rm 1D}}\left[\hat{a}^\dagger(z_n)\s_n^--h.c.\right].
    \end{equation}
\end{subequations}
The system above constitutes the M-B model, which we later use to investigate the possibility of a phase transition in free space. However, before doing so, we show the result of integrating out the field to arrive at an atomic master equation, as in the Cavity model (\ref{eq:Cavity_model}). \\

Starting from Eq.~(\ref{eq:Heisenberg_wQED_fields}), we assume the photon propagation occurs in a time-scale much shorter than the atomic dynamics, as $\Gamma_{\rm 1D}\gg L/c$, where $L$ is the system length. This allows us to disregard the time derivative in Eq. (\ref{eq:Heisenberg_wQED_fields}), which is equivalent to a Markov approximation \cite{Shi2015,Caneva_2015,gross1982}. Integrating over space yields the input-output relation
\begin{equation}
    \label{eq:Input_output_wQED}
    \hat{a}(z)=a_{\rm in} - i\sqrt{\frac{\Gamma_{\rm 1D}}{2c}}\sum_{n=1}^N\s^-_n\theta(z-z_n),
\end{equation}
where $\theta(z-z_n)$ is the Heaviside step function with $\theta(0)=1/2$. In essence, Eq.~(\ref{eq:Input_output_wQED}) shows the field at position $z$ is the coherent sum of the input light $a_{\rm in}$~(taken to be a $c$-number corresponding to a coherent state) and the atomic emission. Consistently, only the atoms at the left of the position $z$ contribute to the total field, as $\hat{a}(z)$ is right propagating.

The fact that the total field driving the atoms in Eq.~(\ref{eq:Heisenberg_wQED_equations}) can be expressed purely in terms of the input field and other atoms suggests that an atom-only description is possible~(see, e.g., Refs.~\cite{Asenjo2017, meystre2007elements} for more details). In addition to the propagating mode of interest, we  assume that the atomic interaction with the remaining free-space modes can be modeled as independent emission $\Gamma$, as in the Cavity model of Sec.~\ref{Sec:Cavity_model}. The resulting master equation for the atomic density matrix $\hat{\rho}$ reads \cite{Giuseppe2022}
\begin{subequations}\label{eq:Chiral_wqed_master_equation}
\begin{equation}
    \dot{\hat{\rho}}=\frac{-i}{\hbar}\left[(\hat{H}_{\rm drive}+\hat{H}_{\rm dd}),\hat{\rho}\right] + \mathcal{L}_{\Downarrow}'[\hat{\rho}] + \mathcal{L}_{\downarrow}[\hat{\rho}],
\end{equation}
\begin{equation}\label{eq:H_dd}
    \hat{H}_{\rm dd} = -\frac{i\hbar\Gamma_{\rm 1D}}{4}\sum_{n>m}\left[\s^{+}_n\s^-_m-\s^{+}_m\s^-_n\right],
\end{equation}
\begin{equation}\label{eq:collective_dissipation_wqed_ME}
        \mathcal{L}_{\Downarrow}'[\hat{\rho}]=\frac{\Gamma_{\rm 1D}}{4}\left[2\hat{S}^-\hat{\rho} \hat{S}^+-\left\{\hat{\rho},\hat{S}^+\hat{S}^-\right\}\right],
\end{equation}
\end{subequations}
where $\hat{H}_{\rm drive}=\hbar \sqrt{c\Gamma_{\rm 1D}/2}\sum_n (a_{\rm in}^*(z_n) \s^-_n+h.c.)$ describes the driving by the input field and $\hat{H}_{\rm dd}$ describes coherent dipole-dipole interactions mediated by photons in the quasi-1D mode.

The master equation (\ref{eq:Chiral_wqed_master_equation}) reveals that the only difference between our free space model and the Cavity model (\ref{eq:Cavity_model}) is the coherent dipole-dipole interaction term $\hat{H}_{\rm dd}$. Note that permutationally invariant dipole-dipole interactions can also appear in the Cavity model, when the atomic and cavity resonance frequencies do not match \cite{Ritsch2013}. However, the $\hat{H}_{\rm dd}$ in Eq.~(\ref{eq:H_dd}) is different in nature, as it depends on the relative atomic positions and cannot be made to vanish by some choice of atomic resonance frequency. Intuitively, this term is necessary to capture pulse propagation within the ensemble, as is allowed in free space, and thus breaks the permutational symmetry of the Cavity (or Dicke) model. \\

We emphasize that the M-B model shown in Eqs.~(\ref{eq:Heisenberg_wQED_equations}) is simply a minimal~(but oft-used) model for light propagation in atomic ensembles. As the emission into the modes not of explicit interest is considered to be independent, it neglects effects associated with wave interference and multiple scattering of light in these modes. For example, this model cannot capture the ``speckled'' pattern of the scattered far-field intensity when the atomic positions are disordered ~\cite{Lagendijk1996,Schilder2017}, nor can it capture Bragg diffraction~\cite{Birkl1995} or the perfect reflection of light from ordered arrays~\cite{Bettles2015, Shahmoon, Rui2020}, which has recently been realized to be a valuable resource for applications~\cite{Manzoni_2018,Bekenstein2020,MorenoCardoner2021}. In dense ensembles, the dipole-dipole interactions that result from atoms experiencing the scattered fields of other atoms can result in significant shifts in the collective atomic energies and decay rates \cite{Asenjo2017, Glicenstein2020, Ferioli2021,Jennewein_2018,Jennewein_2016}, fundamental dephasing mechanisms \cite{ji2023}, or limitations to the ensemble refractive index \cite{Andreoli2021}.

While the M-B model certainly misses the physics above, these effects are generally believed to be negligible in dilute atomic clouds. Our focus remains to describe the propagation of the input field and its interaction with the ensemble, which justifies our quasi-one-dimensional model where the remaining $4\pi$ of modes is treated as independent emission. We also again point to the complementary work of Ref.~\cite{Argawal_2024}, where the three-dimensional nature of the problem is treated in more detail, and which arrives at similar conclusions. 

\subsection{Mean-field steady-state solution}
The M-B equations generally have no exact solution beyond the limit of weak driving~(linear response), so we resort to a mean-field treatment to look for a phase transition in the thermodynamic limit. We include the photonic degrees of freedom back in the discussion and go back to our Eqs.~(\ref{eq:Heisenberg_wQED_equations}) with the addition of the independent emission term $\mathcal{L}_\downarrow[\hat{\rho}]$. To further simplify the equations, we replace the granular ensemble with a macroscopically smooth medium. This is done by substituting the atomic operators with continuous quantum fields satisfying $\sum_{n=1}^N\s^-_n\delta(z-z_n)\approx \frac{N}{L}\s^-(z)$. The resulting equations of motion have the usual M-B form \cite{scully_zubairy_1997}
\begin{subequations}\label{eq:Maxwell_Bloch_continuum}
    \begin{equation}\label{eq:Maxwell_Bloch_continuum_field}
        \partial_z\hat{E}(z)=-i\sqrt{\frac{\Gamma_{\rm 1D}}{2}}\frac{N}{L}\s^{-}(z),
        \end{equation} 
           \begin{equation}\label{eq:Maxwell_Bloch_continuum_atoms1}
         \dot{\s}^{-}(z)=-\frac{\Gamma}{2}\s^{-}(z)+i\sqrt{\frac{\Gamma_{\rm 1D}}{2}}\hat{E}(z)\s^z(z),
    \end{equation}
       \begin{equation}\label{eq:Maxwell_Bloch_continuum_atoms2}
         \dot{\s}^{z}(z)=-\Gamma(\s^{z}(z)+1)+i\sqrt{2\Gamma_{\rm 1D}}(\hat{E}^\dagger(z)\s^{-}(z)-h.c.),
    \end{equation}
\end{subequations}
where the spatiotemporal dependence of the fields is implicit and we have included the effects of local dissipation $\Gamma$. Here, we have also normalized the field $\hat{E}=\sqrt{c}\, \hat{a}$ such that $\langle \hat{E}^\dagger \hat{E}\rangle$ represents the number of photons passing through the plane $z$ per unit of time. In addition, we have neglected the quantum Langevin noise operators that accompany the dissipation in Eqs.~(\ref{eq:Maxwell_Bloch_continuum_atoms1}) and (\ref{eq:Maxwell_Bloch_continuum_atoms2}), as they play no role at the mean-field level. \\

The mean-field solution is obtained by replacing any operator $\hat{O}$ by the sum of their expectation value and fluctuation $\hat{O}=O+\hat{\delta}_O$, and then neglecting $\hat{\delta}_O$ to achieve a solution to lowest order \cite{gross1982, Kasper2023}. Taking $\Gamma\gg \Gamma_{\rm 1D}$ as in realistic experimental scenarios, the steady-state solutions of Eqs.~(\ref{eq:Maxwell_Bloch_continuum_atoms1}) and (\ref{eq:Maxwell_Bloch_continuum_atoms2}) read
\begin{equation}
\sigma^{-}(z)=-\frac{i\sqrt{2\eta\Gamma}E(z)}{\Gamma+4\eta E^2(z)},\quad \sigma^{z}(z)=\frac{-\Gamma}{\Gamma+4\eta E^2(z)},
    \label{eq:steady_state_atoms}
\end{equation}
where $\eta=\Gamma_{\rm 1D}/\Gamma$ is the single-atom cooperativity, which physically represents the ratio of the emission rates into the Gaussian mode and the remaining free space modes. In terms of physical parameters \cite{Tey_2009, Goncalves_2021},
\begin{equation}
    \eta = \frac{3}{8\pi}\frac{\lambda^2}{A},
    \label{eq:cooperativity}
\end{equation}
where $A$ is the mode area of the driving beam and $\lambda$ is the resonant wavelength of the atomic transition. This guess will be refined later in Sec.~IV to account for the specific could geometry and natural emission mode of the ensemble, while we keep a more general description here.\\

Substituting Eq.~(\ref{eq:steady_state_atoms}) into Eq.~(\ref{eq:Maxwell_Bloch_continuum_field}) yields a first-order differential equation with an analytic solution. Imposing the boundary condition $E(0)= E_{\rm in}$, the amplitude of the field within a 1D ensemble spanning $z\in[0,L]$ obeys
\begin{multline}
           E(z)=E_{\rm in}\ \text{exp}\Bigg{[}\frac{D}{2} \left(s-\frac{z}{L}\right)-\\
       -\frac{1}{2}W\left(\text{log}(D \   s)+D \left(s-\frac{z}{L}\right)\right)\bigg{]}.
        \label{eq:steady_state_2}
\end{multline}
Here, $W(x)$ is the Wright Omega function \cite{Wright_omega_function, Kasper2023}, whose specific form will not be relevant. The only important property is that $W(x)$ exhibits linear behavior for large positive values of $x\gg 1$ and exponential behavior for negative ones $x\ll -1$. The solution in Eq.~(\ref{eq:steady_state_2}) is mathematically equivalent to the problem of light propagation through a classical saturable absorber \cite{Veyron2021, Kim_2020}.

Equation (\ref{eq:steady_state_2}) suggests that the mean-field behavior of our system is entirely determined by two experimentally relevant parameters: the optical depth $D=2\eta N$ and the saturation parameter $s= 2|E_{\rm in}|^2/ N \Gamma$. The former characterizes the degree of exponential attenuation of a weak field in transmission, while the latter is the ratio between the input photon flux and the photon flux radiated by a completely saturated ensemble of $N$ atoms into $4\pi$. (Note that a very strongly driven atom becomes completely mixed in steady-state, and thus $(N/2)\Gamma$ is the maximum rate of photons that could be scattered by the ensemble.) 

For $s\ll1$, the system is weakly driven and the input field is exponentially attenuated across the ensemble. One can explicitly see this from the transmittance in the M-B model, which reads
\begin{equation}
    T(s, D)=\frac{ E^2(L)}{ E^2(0)}.
    \label{eq:MB_Transmission_coherent}
\end{equation}
For small $s$, Eqs.~(\ref{eq:MB_Transmission_coherent}) and (\ref{eq:steady_state_2}) lead to  $T\approx e^{-D}$, consistently with the definition of optical depth. In the opposite regime, where $s\gg1$, the system is driven strongly and the atomic response saturates ($\sigma^-,\sigma^z\approx 0$). Physically, the input field greatly exceeds the atomic emission and the field $E(z)\approx E_{\rm in}$ remains roughly constant throughout the ensemble, yielding $T\approx 1$.

\subsection{Existence of phase transition?}\label{Sec:Phase_tranisition}

\begin{figure}[t!]
    \centering
    \includegraphics[width=0.98\linewidth]{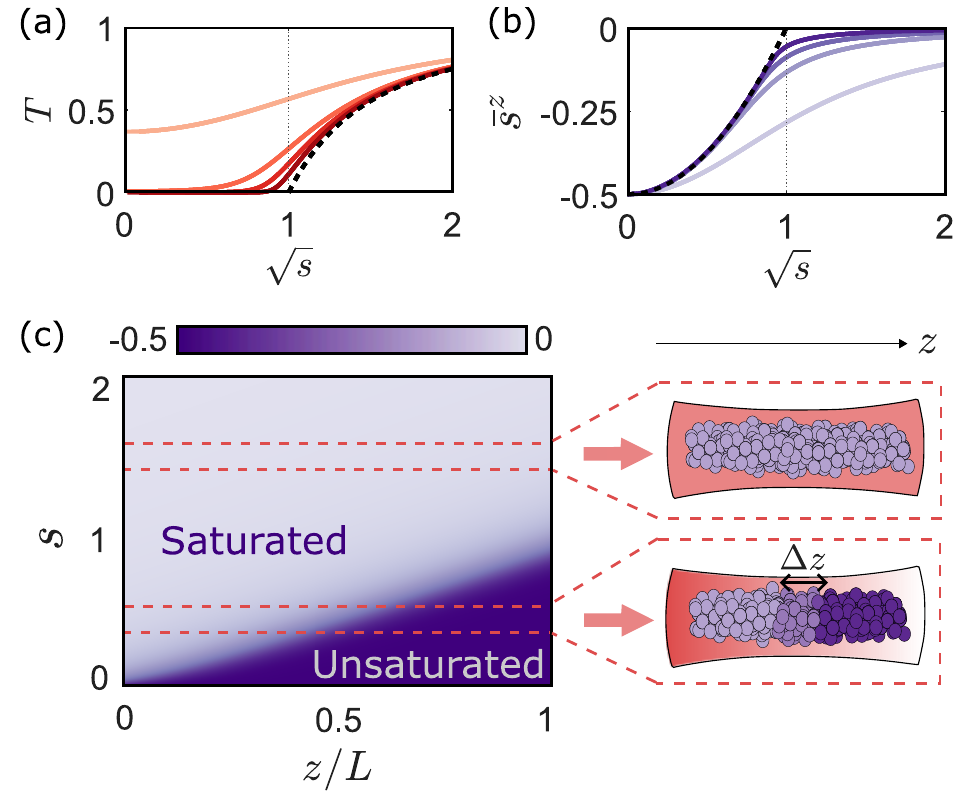}
    \caption{(a,b) Transmission $T$ and averaged magnetization $\bar{s}^z$ in the M-B model as a function of the square root of the saturation parameter $\sqrt{s}$ for different optical depths $D=\{1,5,10,20\}$ (light to darker colors). We also plot the respective thermodynamic limits from Eqs. (\ref{eq:Infinite_D_Transmission}) and (\ref{eq:Infinite_D_Magnetization}) in dashed black. Both observables exhibit a sharp transition at $s=1$. (c) Local magnetization $s^z(z)$ as a function of position $z$ and saturation parameter $s$ for $D=20$. In the magnetized phase ($s<1$), the system splits into saturated and unsaturated regions at $z\approx z_c$. Here, $\Delta z$ is the length of the boundary between the two regimes. In the unmagnetized phase ($s>1$), the system is completely saturated.} 
    \label{fig:MB_phase_separation}
\end{figure}

Phase transitions are characterized by the non-analytic behavior of observables (or their derivatives) in the thermodynamic limit upon the change of a parameter. For example, in the Dicke model, the derivative of $S_z$ in Eq.~(\ref{eq:Mean_field_magnetization_Dicke}) becomes non-analytic at $\beta=1$. In the M-B model, the thermodynamic limit corresponds to $D\rightarrow\infty$ while maintaining constant $N/AL$ (fixed density). From Eq.~(\ref{eq:MB_Transmission_coherent}), we verify numerically that the transmission for large optical depth satisfies
\begin{equation}
\label{eq:Infinite_D_Transmission}
     T_\infty(s) \approx
    \begin{cases}
      0 & \text{for  $s<1$,}\\
      1-\frac{1}{s} & \text{for $s\geq1$},\\
    \end{cases}
\end{equation}
to leading order in $1/s$. We clearly observe two distinct regimes. In the first regime ($s<1$), the system is magnetized, and the infinite optical depth leads to zero transmission. In other words, the input photons are scattered into free space before they reach the end of the ensemble. In the second regime $(s>1)$, the influx of photons surpasses the maximum atomic emission rate, resulting in atomic saturation and a non-zero transmission. We compare the transmission calculated with Eq.~(\ref{eq:MB_Transmission_coherent}) and the prediction from Eq.~(\ref{eq:Infinite_D_Transmission}) in Fig.~\ref{fig:MB_phase_separation}a as a function of $s$, where we observe a convergence for increasing system sizes. Here, we plot the observables in terms of $\sqrt{s}\propto E_{\rm in}$ to compare later with the Dicke model.

We can also calculate the averaged magnetization $\bar{s}^z=\frac{1}{2L}\int \sigma^z(z)\text{d}z$ in the ensemble by substituting Eq.~(\ref{eq:steady_state_2}) into Eq.~(\ref{eq:steady_state_atoms}). In the thermodynamic limit, one has
\begin{equation}
     \bar{s}^z_\infty(s) \approx
    \begin{cases}
     \frac{s-1}{2} & \text{for  $s<1$,}\\
      0 & \text{for $s\geq1$}.\\
    \end{cases}    
    \label{eq:Infinite_D_Magnetization}
\end{equation}
We again observe two distinct regimes as a function of $s$, where the average magnetization is non-zero for $s<1$ and zero for $s>1$. This is illustrated in Fig.~\ref{fig:MB_phase_separation}b, where we notice a striking similarity to the Dicke phase transition in Fig.~\ref{fig:Magnetization_Dicke_cavity}a. \\

Upon closer inspection, one notices significant differences with the results of the Cavity model. In the magnetized phase, the spatial distribution of the magnetization $s^z(z)=\sigma^{z}(z)/2$ is extremely inhomogeneous. Figure \ref{fig:MB_phase_separation}c provides a visual representation of a system with large optical depth $D=20$, showing that for $s<1$ the side of the ensemble where light enters is saturated and has zero magnetization. However, these atoms scatter all of the incident light into free space, leaving the atoms on the output side un-excited.

Indeed, the behavior in Eq.~(\ref{eq:Infinite_D_Magnetization}) can be easily reproduced by splitting the ensemble into two regions: fully saturated $(\sigma^z(z)=0)$ and not illuminated $(\sigma^z(z)=-1)$. Defining the ``critical" position $z_c$ between these two regions as the position where the variable in the function $W(x)$ of Eq.~(\ref{eq:steady_state_2}) changes sign, i.e. $z_c\equiv sL$, the average magnetization reads
\begin{equation}
    \bar{s}_\infty^z=\frac{1}{2L}\int_{0}^{z_c}\sigma^z(z)\text{d}z+\frac{1}{2L}\int_{z_c}^{L}\sigma^z(z)\text{d}z=\frac{s-1}{2},
\end{equation}
which is exactly Eq.~(\ref{eq:Infinite_D_Magnetization}). This is in stark contrast with the magnetized phase existing in the Cavity and Dicke models, where each individual atom had the same magnetization.

We note that although the simple model above predicts well the average magnetization, the position-dependent magnetization $s^z(z)$ is actually a smooth function, as illustrated in Fig. \ref{fig:MB_phase_separation}c. The transition between the magnetized and saturated regions of the system occurs not at a single point $z_c$ but over a length scale $\Delta z\propto L/D$. However, in the thermodynamic limit, this length scale becomes infinitesimally small compared to the total length $\Delta z/L\propto 1/D\rightarrow 0$, and leads to the non-analytic behavior of the average magnetization. 

Furthermore, even in the global observables like $\hat{S}_z$ and $T$ that exhibit non-analytic behavior, the ``critical value" $s=2E_{\rm in}^2/N\Gamma=1$ depends on field intensity, rather than field amplitude as in the Cavity model. From these arguments, it thus appears that the physics of the ``transition'' is fundamentally different from the ones of the Dicke and Cavity models. For lack of better terminology, one might say that the free-space ensemble exhibits a \textit{phase separation}, with distinct regions of the system exhibiting different behaviors, rather than a phase transition. 

\subsection{Mapping to Cavity QED}\label{Sec:Mapping_cavity_and_MB}
Above, we have argued that the behavior of a free-space ensemble in the thermodynamic limit (high $D$) is fundamentally different than in the case of a cavity. This is due to propagation effects in the former system, which allows for spatial inhomogeneity in local observables. On the other hand, for low $D$, the atoms cannot have a significant effect on the input field. This causes the atoms to respond nearly homogeneously, and in this regime, the system behavior can resemble cavity QED, as we now argue in more detail.

Let us start from the M-B model in Eqs.~(\ref{eq:Maxwell_Bloch_continuum}). We first integrate Eq.~(\ref{eq:Maxwell_Bloch_continuum_field}) with boundary conditions $E(0)=E_{\rm in}$ and substitute the result in Eq.~(\ref{eq:Maxwell_Bloch_continuum_atoms1}). Then, we use the relation $\Omega = \sqrt{2\Gamma_{\rm 1D}}E_{\rm in}$ to rewrite the input field in the units of the Cavity model, and perform a mean-field approximation to decorrelate the operators. Finally, we assume that at low $D$ the ensemble is spatially homogeneous, which allows us to approximate $\langle \s^{\alpha}_i\rangle\approx\langle \s^{\alpha}\rangle$. The resulting mean-field Heisenberg equation for $\hat{S}^-$ looks remarkably similar to the one of the Cavity model for the same assumptions, where 
\begin{subequations}\label{eq:comparing_models}
    \begin{equation}
       \langle\dot{S}_{-}^{\rm Cav}\rangle\approx-\frac{\Gamma N}{2}\langle \s^-\rangle+\frac{i\Omega N}{2}\langle \s^z\rangle+\frac{\Gamma_{\rm 1D}N^2}{2} \langle \s^z\rangle \langle \s^-\rangle,
    \end{equation}
        \begin{equation}
       \langle\dot{S}_{-}^{\rm MB}\rangle\approx-\frac{\Gamma N}{2}\langle \s^-\rangle+\frac{i\Omega N}{2}\langle \s^z\rangle+\frac{\Gamma_{\rm 1D}N^2}{4} \langle \s^z\rangle \langle \s^-\rangle.
    \end{equation}
\end{subequations}
The only difference is a factor $1/2$, which arises from the fact that within the M-B model, the radiation of a given atom into the quasi-1D mode can only affect atoms to the right~($\sim N^2/2$ combinations instead of $\sim N^2$), a so-called ``chiral'' interaction.\\

\begin{figure}[t!]
    \centering
    \includegraphics[width=0.9\linewidth]{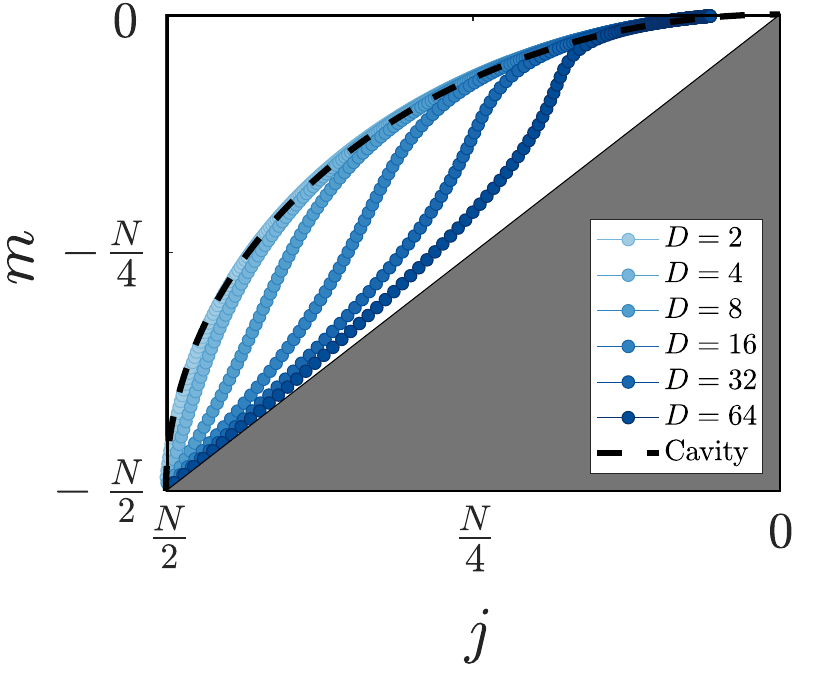}
    \vspace{-1em}
    \caption{Steady-state relationship between magnetization and total angular momentum in the M-B model. The points correspond to the estimated quantum numbers $m=\langle \hat{S}_z\rangle$ and $j(j+1)=\langle \hat{S}^2\rangle$ from a numerical simulation of the M-B model for different optical depths $D$ (light to dark blue). For a fixed optical depth, the different points are obtained by varying the driving strength. We also plot the steady-state universal relation (\ref{eq:relation_from_lateral_balance}) from the Cavity model with black dashed lines. The models agree when the optical depth is low, and/or the driving strength is sufficiently large.}    \label{fig:Maxwell_Bloch_S_Sz_trajectory}
\end{figure}

To further support this mapping, we use Eqs.~(\ref{eq:steady_state_atoms}) and (\ref{eq:steady_state_2}) to calculate the steady-state magnetization $\langle \hat{S}_z\rangle$ and total angular momentum $\langle \hat{S}^2\rangle$ in the M-B model for varying optical depths $D$ and driving strengths. As we illustrate in Fig.~\ref{fig:Maxwell_Bloch_S_Sz_trajectory}, the results from the M-B model obey the same universal relation (\ref{eq:relation_from_lateral_balance}) from the Cavity model at low $D$. To calculate the expectation values of observables, such as $\hat{S}^2$, we used a product state ansatz, which we discuss in more detail in the next section.
  
The fact that an ensemble becomes homogeneous at low $D$ has important implications for the dipole-dipole interactions. Starting from the effective Hamiltonian $\hat{H}_{\rm dd}$ in Eq.~(\ref{eq:H_dd}), one can see that the combination of mean-field and spatial homogeneity renders the coherent part of the dipole-dipole interactions negligible. Explicitly, 
    \begin{equation}
        \langle \hat{H}_{\rm dd}\rangle \approx \frac{\Gamma_{\rm 1D}N^2}{8}(\langle \s^+\rangle \langle \s^-\rangle -\langle \s^+\rangle \langle \s^-\rangle )=0.
    \end{equation}
This again illustrates the role of the dipole-dipole interactions in encoding field propagation (and thus spatially inhomogeneous response) in the ensemble.\\

In addition to the low $D$ limit, the Cavity and M-B models converge for sufficiently large driving strengths. In this limit, the atoms become saturated, the atomic coherences $\langle \hat{\sigma}^{\pm}_n\rangle$ vanish and the chiral dipole-dipole interactions become negligible $\langle \hat{H}_{\rm dd}\rangle\approx 0$. This convergence can also be appreciated in Fig.~\ref{fig:Maxwell_Bloch_S_Sz_trajectory}, where the Cavity and M-B curves coincide for large enough driving (small enough $m$) regardless of the optical depth. In Appendix B, we provide a complementary analysis based on how $\hat{H}_{\rm dd}$ acts on the angular momentum states in the strong driving limit.

\section{Modeling of experimental data}

We now reproduce some key experimental results from Ref.~\cite{Ferioli2023} using the M-B model. Instead of the natural parameters $D$ and $s$ of the M-B model, the experiment had access to atom numbers $N$ and Rabi frequencies $\Omega$ \cite{Ferioli2023}. Thus, to relate these quantities, we need to establish first a value for the cooperativity $\eta$ of the experiment.

To do this, we need to point out a subtlety that arises when modeling a 3D ensemble by the 1D M-B model. Within the 1D M-B model, the cooperativity $\eta$ actually has two physical meanings. First, from Eq.~(\ref{eq:cooperativity}), we see that $\eta$ defines the emission rate of a single atom into the 1D mode of the M-B model (the Gaussian mode of the driving beam). However, from Eq.~(\ref{eq:Chiral_wqed_master_equation}), $\eta$ also characterizes the ratio of independent to collective dissipation strengths of a collective atomic spin wave excitation. Within the 1D M-B model, where there is no explicit concept of transverse mode shapes, these two meanings naturally coincide. 

However, in a 3D geometry such as an elongated ensemble, a collective spin wave excitation will have an emission pattern or transverse mode structure that is not necessarily the same as the input mode \cite{gross1982}. If the collective atomic excitation is constructively emitting into a significantly larger solid angle~(as is the case of the experiment), then the $\eta$ as calculated from Eq.~(\ref{eq:cooperativity}), where only the contribution into the solid angle of the driving mode is considered, will significantly underestimate the collective dissipation rate.

In order to arrive at a more realistic model for $\eta_{\rm eff}$ that accounts for this effect, we consider the total power $P$ emitted by a cloud of $N$ atoms into a solid angle $\Delta \Theta$ in free space, \cite{Allen_1975}
\begin{equation}    P=\int_{\Delta\Theta}\text{d}\Omega_\textbf{k}\ I_N(\textbf{k}),
    \label{eq:total_power_atomic_emission}
\end{equation}
where $I_N(\textbf{k})$ is the intensity emitted in the direction of the wavevector $\textbf{k}$. For a single atom, and a full solid angle of $\Delta\Theta=4\pi$, Eq.~(\ref{eq:total_power_atomic_emission}) can be written in terms of the free-space emission rate $\Gamma$ as
\begin{equation}
   P=\int_{ 4\pi}\text{d}\Omega_\textbf{k} I_1(\textbf{k})\equiv\Gamma\langle \s^+\s^-\rangle,
\end{equation}
where $I_1(\textbf{k})$ is the far-field radiation pattern of a single atom in free space. This normalization is equivalent to that of the M-B model, where $P$ represents the number of photons emitted per unit of time. 

\begin{figure}[t!]
    \centering
    \includegraphics[width=0.9\linewidth]{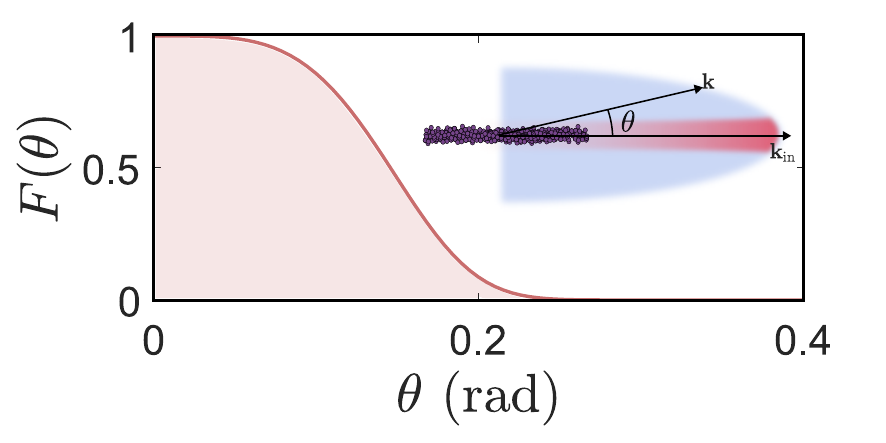}
    \vspace{-1em}
    \caption{Form factor $F(\textbf{k},\textbf{k}_{\rm in})$ as a function of the angle $\theta$ between $\textbf{k}$ and the input field $\textbf{k}_{\rm in}$, and calculated for typical ensemble sizes of the experiment of Ref.~\cite{Ferioli2023}. The ensemble exhibits coherent emission for small angles $\theta\lesssim 0.2$ rad, while emission is mainly incoherent for
    larger $\theta$. Inset: Sketch of the incoherent (blue) and coherent (red) components of the light emitted by an elongated ensemble.}
    \label{fig:Components_atomic_emission}
\end{figure}

We take a simple model for the total power, where we assume the ensemble has a spatially homogeneous response $\langle\s_n^-\rangle \approx \langle \s^-\rangle e^{ik_{\rm in}z_n}$, where $k_{\rm in}$ is the input field wavevector. For $N\gg 1$, the intensity $I_N(\textbf{k})$ emitted in the direction $\textbf{k}$ can be separated into coherent and incoherent components as \cite{Ferioli2023,TANJISUZUKI2011201, Robicheaux_2017}
\begin{subequations}\label{eq:Components_intensity_k}
   \begin{equation}
    I_{\rm coh}(\textbf{k})=I_1(\textbf{k})N^2|\langle \s^-\rangle|^2F(\textbf{k},\textbf{k}_{\rm in}),
\end{equation}
\begin{equation}
    I_{\rm incoh}(\textbf{k})=I_1(\textbf{k})N\langle \s^+\s^-\rangle,
\end{equation}
\end{subequations}
with only the former having a well-defined phase relation with the input field. Here, $F(\textbf{k},\textbf{k}_{\rm in})$ is a form factor that depends on the angle $\theta$ between $\textbf{k}$ and $\textbf{k}_{\rm in}$ and the distribution of atomic positions. Assuming the atomic positions $z_n$ lie on the optical axis, it reads
\begin{equation}
   F(\textbf{k},\textbf{k}_{\rm in})=\frac{1}{N^2}\sum_{n\not=m}e^{ik_{\rm in}(1-\cos\theta)(z_n-z_m)}.
   \label{eq:form_factor}
\end{equation}
Physically, $F(\textbf{k},\textbf{k}_{\rm in})$ captures the angle-dependent interference between atoms in the coherent atomic emission. The key observation is that for random $z_n$, Eq.~(\ref{eq:form_factor}) predicts constructive interference only within a small solid angle centered around $\textbf{k}_{\rm in}$. Indeed, assuming $z_n$ follow a Gaussian distribution $\rho(z;\sigma_L)$ with variance $\sigma_L^2$ and zero mean, Eq.~(\ref{eq:form_factor}) in the continuous limit becomes \cite{Ferioli2023}
\begin{multline}
       F(\theta)\approx \left|\int_{-\infty}^{\infty} \text{d}z\ \rho(z;\sigma_L)\ e^{ik_{\rm in}(1-\cos\theta)z} \right|^2=\\= \text{exp}[-\sigma_L^2k_{\rm in}^2(1-\cos\theta)^2].
   \label{eq:form_factor_continous}
\end{multline}
Figure (\ref{fig:Components_atomic_emission}) shows Eq.~(\ref{eq:form_factor_continous}) as a function of the angle $\theta$, where we see that $F(\theta)$ vanishes for angles larger than $\theta\lesssim0.2$ rad. Here, we have considered a standard deviation of $\sigma_L
\approx 12.5\lambda$. We note that estimating $\sigma_L$ experimentally is quite challenging, and the value chosen here is roughly half the measured axial size $l_{\rm ax}\sim 20-25\lambda$ in Ref.~\cite{Ferioli2023}, consistently with the discussion surrounding the Appendix C of Ref.~\cite{Ferioli2023}.\\

We then estimate the effective single-atom cooperativity $\eta_{\rm eff}$ as the fraction of coherently-scattered light in the forward direction $(k_z>0)$, in the weak driving limit $\langle \hat{\sigma}^+\hat{\sigma}^-\rangle \approx |\langle \hat{\sigma}^-\rangle|^2$,
\begin{equation}
   \eta_{\rm eff} =\frac{\int_{k_z>0}\text{d}\Omega_\textbf{k}F(\textbf{k},\textbf{k}_{\rm in})I_1(\textbf{k})}{\int_{k_z>0}\text{d}\Omega_\textbf{k}I_1(\textbf{k})}.
    \label{eq:definition_eta_effective}
\end{equation}
Intuitively, $\eta_{\rm eff}=1$ when atoms constructively interfere \textit{everywhere} in the forward direction, like in the traditional Dicke scenario where all atoms are localized at the same point ($\sigma_L=0$) \cite{Dicke54}. On the other hand, $\eta_{\rm eff}=0$ when atoms can have arbitrary relative phases and do not constructively interfere. In the case of the experiment \cite{Ferioli2023}, where atoms were circularly polarized $I_1(\textbf{k})=3\Gamma(1+\sin^2\theta\cos^2\phi)/16\pi$ and $\sigma_L\approx12.5\lambda$, Eq.~(\ref{eq:definition_eta_effective}) yields an effective cooperativity of $\eta_{\rm eff}\approx 0.01$. We note this is significantly larger than the value of $\eta\sim 0.002$ if one uses Eq.~(\ref{eq:cooperativity}) for the driving field in Ref.~\cite{Ferioli2023}, describing the emission efficiency of a single atom into the input mode. 

Finally, let us detail how we calculate the parameters of the M-B model. First, the optical depth $D$ is obtained from the atom numbers $N$ measured in Ref.~\cite{Ferioli2023}, using the relation $D=2\eta_{\rm eff}N$. For the saturation parameter $s$, we use the relation
\begin{equation}
    \beta=\frac{2\Omega}{N\eta_{\rm eff}\Gamma}\equiv\sqrt{\frac{8s}{D}},
    \label{eq:relation}
\end{equation}
where we substitute the calculated value of $D$ along with the measured values of $\Omega$ and $N$. This expression also relates the parameter $\beta$ that dictates the Dicke phase transition with the M-B model.

\subsection{Transmission and magnetization}

\begin{figure}[t!]
    \centering
    \includegraphics[width=0.98\linewidth]{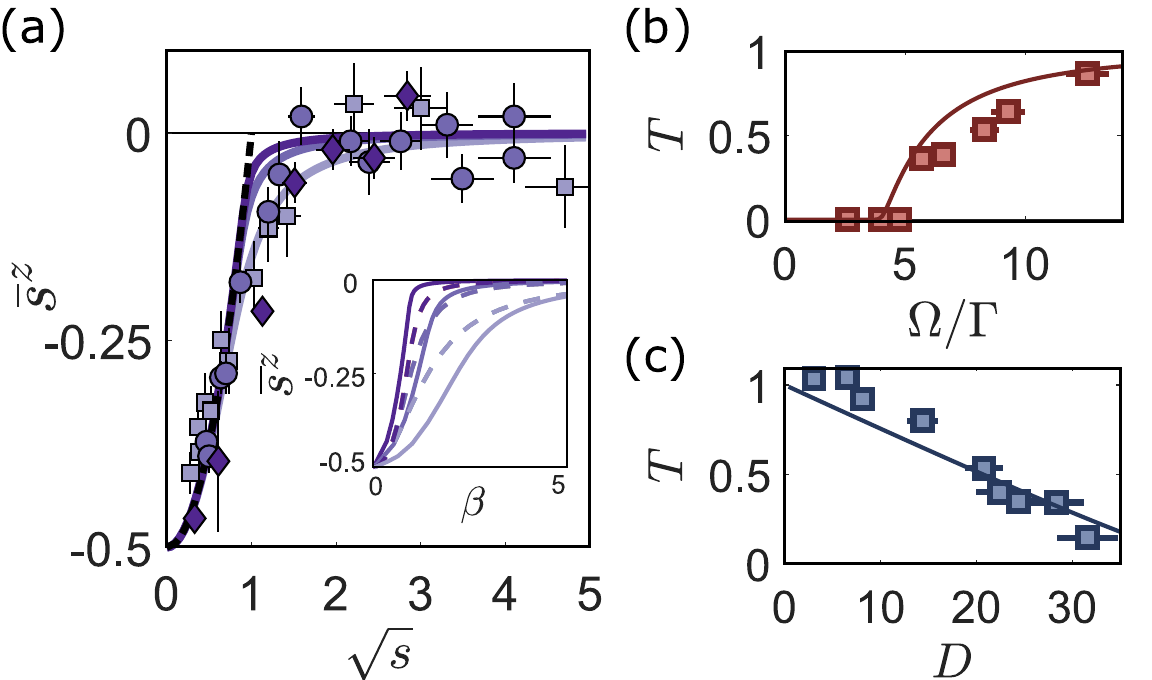}
    \caption{(a) Averaged steady-state magnetization $\bar{s}^z$ as a function of $\sqrt{s}\propto \Omega$ for $D=[4,12,32]$ (light to dark purple). The points are experimental data from Ref.~\cite{Ferioli2023}, with $D=[4,12,32]$ corresponding to the squares, circles, and diamonds, respectively. The black-dashed curve is the large $D$ limit from Eq.~(\ref{eq:Infinite_D_Magnetization}). Inset: Same averaged magnetization, but now as a function of the $\beta$ as defined in Ref.~\cite{Ferioli2023}. Solid lines are obtained with the M-B model, while we used dashed lines for the Dicke model. (b) Steady-state transmittance $T$ as a function of driving strength $\Omega$ and fixed optical depth $D$. The red squares are experimental values of $T\approx (\Omega_{\rm eff}/\Omega)^2$ from Ref.~\cite{Ferioli2023}, and the solid line is the theoretical prediction of Eq.~(\ref{eq:MB_Transmission_coherent}), with the optical depth of $D=36$ inferred from experimental parameters. Phase separation appears at $\Omega\approx 5\Gamma$. (c) Same as (b) but for fixed $\Omega=4.5\Gamma$ and varying $D$. }
    \label{fig:Magnetization_and_transmission}
\end{figure}

We start by calculating the average magnetization $\bar{s}^z=\langle \hat{S}_z\rangle/N$ for various optical depths $D$. In Fig.~\ref{fig:Magnetization_and_transmission}a, we plot the prediction from Eqs.~(\ref{eq:steady_state_atoms}) and (\ref{eq:steady_state_2}) as a function of $\sqrt{s}$, in solid lines. The results agree with both the thermodynamic limit from Eq.~(\ref{eq:Infinite_D_Magnetization}), in dashed-black, and the measurements~(points in the figure) in Ref.~\cite{Ferioli2023}. Again, we plot the average magnetization in terms of $\sqrt{s}$ because it is proportional to the field amplitude, like $\beta$ in the Cavity model. 

In the inset of Fig.~\ref{fig:Magnetization_and_transmission}a, we compare the average steady-state magnetization in both the M-B and Dicke models~(solid and dashed curves, respectively) as a function of $\beta$ for the same number of atoms. We see a remarkable similarity between the predictions of both models in the parameter regime studied by Ref.~\cite{Ferioli2023}. We also note that, for larger $D$, the transition point in the M-B model approaches $\beta_c\sim 0$, instead of converging to $\beta_c=1$ as in the Dicke model.\\

We now calculate the transmittance $T$ using Eq.~(\ref{eq:MB_Transmission_coherent}). In Fig.~\ref{fig:Magnetization_and_transmission}b, we plot $T$ as a function of driving amplitude $\Omega$ and for an optical depth of $D=36$. Importantly, $T$ was not measured in Ref.~\cite{Ferioli2023}, and thus we cannot directly compare theory and experiment.

Nevertheless, one can infer an approximated value for $T$ from the measurements of $\Omega_{\rm eff}$ in Ref.~\cite{Ferioli2023}. There, $\Omega_{\rm eff}$ is the oscillation frequency of the averaged magnetization $\bar{s}^z(t)$ at early times ($t\lesssim 5\Gamma$). Assuming the ensemble is homogeneous enough, it is given by the total field inside the ensemble $\Omega_{\rm eff}\approx\Omega-i\Gamma_{\rm 1D}\langle\hat{S}^z\rangle$, which is our Eq.~(\ref{eq:Input_output_wQED}). In Fig.~\ref{fig:Magnetization_and_transmission}b, we plot this frequency normalized by the input $\Omega$, yielding an approximate transmission $T\approx (\Omega_{\rm eff}/\Omega)^2$.

Comparing the transmittance from the M-B model and the one inferred from the measurements of $\Omega_{\rm eff}$, we observe an overall good agreement, including the sharp transition around $\Omega\approx 5\Gamma$ (see Eq.~(\ref{eq:Infinite_D_Transmission})). 
We also observe a good agreement between the M-B model prediction and the experiment when studying the transmittance as a function of $D\propto N$ and for a fixed driving strength $\Omega$, as shown in Fig.~\ref{fig:Magnetization_and_transmission}c.

\subsection{Atomic emission}
\begin{figure}[t!]
    \centering
    \includegraphics[width=0.9\linewidth]{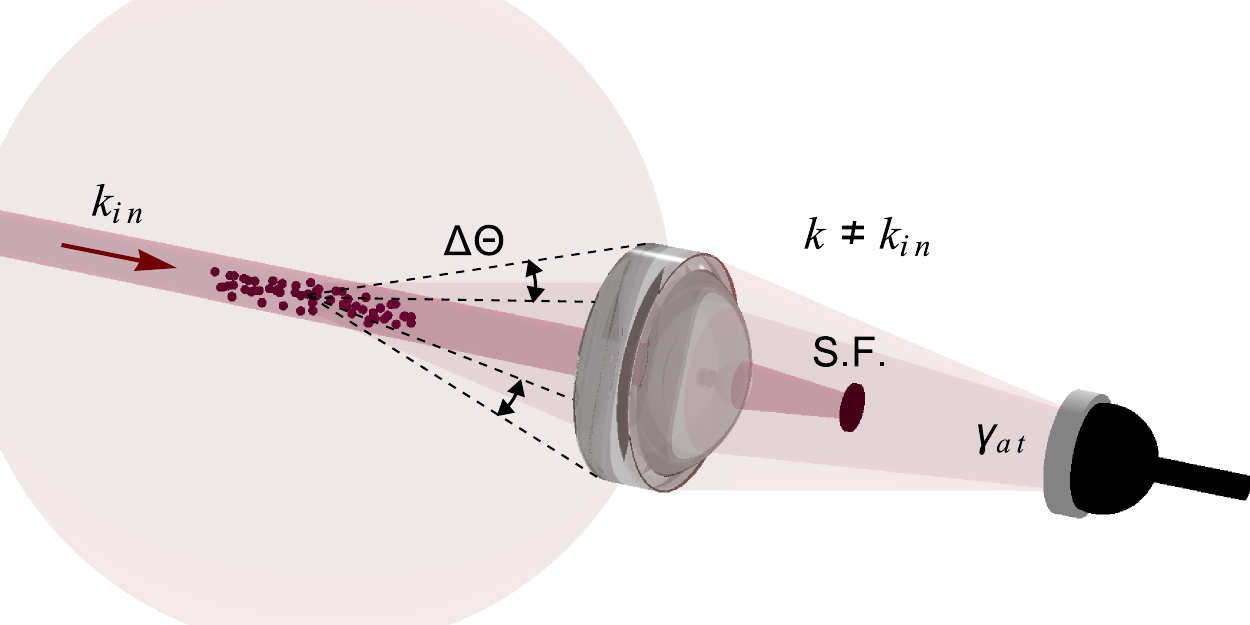}
    \caption{Sketch of the experimental setup in Ref.~\cite{Ferioli2023} for the detection of the atomic field intensity $\gamma_{\rm at}$. A spatial filter (SF) blocks the input field component (dark red) by eliminating modes with $\textbf{k}\approx \textbf{k}_{\rm in}$. The measured atomic emission comprises modes slightly out of the optical axis $\textbf{k}\not\approx \textbf{k}_{\rm in}$ that still fall within the numerical aperture of the lens, defining the solid angle $\Delta\Theta$. The measured $\gamma_{\rm at}$ thus contains both coherent (light red) and incoherent (lighter red) atomic emission.}
    \label{fig:Setup_measurement_GammaSR}
\end{figure}

We now discuss the intensity of the light emitted by the atomic cloud in the forward (transmitted) direction, $\gamma_{\rm at}$. In the M-B model, this is given by the atomic component of the input-output Eq.~(\ref{eq:Input_output_wQED}) at the end of the ensemble ($z>L$). After normalizing $\hat{E}=\sqrt{c}\hat{a}$ so that the emitted light has units of photon number per unit time, one has
\begin{equation}
   \gamma_{\rm at}= \frac{\Gamma_{\rm 1D}}{2}\langle \hat{S}^+\hat{S}^-\rangle.
    \label{eq:general_form_atomic_emission}
\end{equation}
Our theoretical calculation cannot be directly compared with the experimental measurement. This is due to a technicality of the setup used to measure this quantity, which is illustrated in Figure \ref{fig:Setup_measurement_GammaSR}a~\cite{Ferioli2023}. Since the total forward field consists of both input light and atomic emission, a spatial filter (SF) is inserted to block the input Gaussian component from detection. Inevitably, this filter also blocks the atomic emission within the same solid angle covered by the Gaussian mode. Thus, the measured $\gamma_{at}$ corresponds to the atomic radiation emitted slightly out of the optical axis, yet still falling within the numerical aperture of the lens. We will later discuss how to approximately account for this filter in our M-B theory. For completeness, though, we first present the predictions for $\gamma_{\rm at}$ within the ideal M-B model. 

\subsubsection{Product state ansatz}
To calculate quantities like atomic emission $\gamma_{\rm at}$ or second-order photon correlations $g^{(2)}(0)$ within the M-B model, it is convenient to undo the continuous limit transformation $\s_n\leftrightarrow\frac{N}{L}\s(z_n)$ on the operators in Eq.~(\ref{eq:Maxwell_Bloch_continuum}) and return to a discrete atom picture. The mean-field approximation used in the previous sections is equivalent to assuming that the atomic density matrix $\hat{\rho}$ is in the product state \cite{Kasper2023}
\begin{equation}
    \hat{\rho} = \bigotimes_{n=1}^N\hat{\rho}_n,\qquad  \hat{\rho}_n=\frac{1}{2}\left(\begin{array}{cc}
        1-\langle \s^{z}_n\rangle & 2\langle \s_n^-\rangle \\
        2\langle \s_n^-\rangle^* & 1+\langle \s^{z}_n\rangle
    \end{array}\right).
      \label{eq:product_state_ansatz}
\end{equation}
Here, $\hat{\rho}_n$ is a local density matrix constructed from the mean-field result of the M-B model at position $z_n=nL/N$ with integer $n\in[1,N]$. As an aside, applying the mean-field approximation $\hat{E}\rightarrow \langle\hat{E}\rangle$ in Eq.~(17) also assumes that the atoms only interact with the {\it coherent} part of the total field. 

Substituting the product state ansatz from Eq.~(\ref{eq:product_state_ansatz}), the atomic emission in Eq.~(\ref{eq:general_form_atomic_emission}) becomes
\begin{equation}
    \gamma_{\rm at}=\frac{\Gamma_{\rm 1D}}{2}\sum_{n=1}^N\frac{\langle \s^{z}_n\rangle+1}{2}+\frac{\Gamma_{\rm 1D}}{2}\sum_{n\not=m}\langle \s^{+}_n\rangle \langle \s^{-}_m\rangle.
    \label{eq:Atomic_emission}
\end{equation}
Equation (\ref{eq:Atomic_emission}) represents the total atomic emission into the 1D mode. The first and second terms are also referred to as incoherent and coherent components of atomic emission, as characterized by the absence~(presence) of a well-defined phase relationship with the input field, analogous to our definitions in Eqs.~(\ref{eq:Components_intensity_k}).

\subsubsection{Steady-state atomic emission}
In the following, we describe the properties of the steady-state atomic emission $\gamma_{\rm at}^{\rm st}$ in Eq.~(\ref{eq:Atomic_emission}), while we discuss its early-time dynamics in Appendix A. Figure \ref{fig:superradiant_emission}a shows $\gamma_{\rm at}^{\rm st}$ as a function of the saturation parameter $s$, calculated with Eqs.~(\ref{eq:steady_state_atoms}), (\ref{eq:steady_state_2}) and (\ref{eq:Atomic_emission}) for different optical depths. In the magnetized regime ($s\ll1$), the phase separation takes place, and only the region with illuminated atoms can radiate, such that $\gamma_{\rm at}^{\rm st}\propto \Gamma Nz_c/L\propto \Gamma Ds \propto|E_{\rm in}|^2$ independently of system size. Conversely, deep in the unmagnetized regime ($s\gg1$), atomic coherence vanishes and $\gamma_{\rm at}^{\rm st}=\Gamma_{\rm 1D} N/4$ like saturated independent atoms, scaling linearly with the number of atoms. 

Between these two extremes, the steady-state atomic emission features a large maximum around $s\sim1$. As shown in the inset of Fig.~\ref{fig:superradiant_emission}a, these maxima seem to scale quadratically with system size for small optical depth, a feature reminiscent of the $N^2$ superradiant scaling of the Dicke model. However, two key differences arise. First, it only appears for $s\sim1$. Unlike the Dicke model, where the conservation of angular momentum preserves superradiance at arbitrarily strong drivings, the free-space ensemble lacks any symmetry and becomes completely saturated for sufficiently large $s\gg 1$, which in this limit ultimately yields $\gamma_{\rm at}^{\rm st}\propto N$. Second, while superradiance in the Dicke model prevails for arbitrarily large systems, the phase separation for large $D$ splits the free-space ensemble into saturated and unsaturated regions (neither of which being superradiant), and destroys the $\sim N^2$ scaling.\\


Let us use the Cavity model to elucidate the origin of the $N^2$ scaling at small optical depths, given independent emission. Similarly to Eq.~(\ref{eq:rate_in}), we calculate the total decay rate $\Gamma^{\rm out}_{m,j}$ from $|j,m\rangle$ to any other state $|j',m-1\rangle$ \cite{Shammah2018}, and rewrite the result in terms of the fraction of excited population $n=m/N+0.5$, using Eq.~($\ref{eq:relation_from_lateral_balance}$) to relate $j$ to $n$. We obtain
\begin{equation}\label{eq:rate_out}
    \Gamma^{\rm out}_{j,m}(n)\approx \Gamma_{\rm 1D}N^2\left(n-2n^2\right)+\Gamma N n,\quad n\leq \frac{1}{2},
    \end{equation}
where we assumed $N\gg1$ and a non-zero $n$. The first term represents the collective emission into the cavity mode, while the second captures the atomic emission into free space. Equation (\ref{eq:rate_out}) exhibits superradiant scaling $\sim N^2$ at its maximum in $n=1/4$, while it becomes linear with $N$ at saturation ($n=1/2$).

From our discussion in Sec.~\ref{Sec:Mapping_cavity_and_MB}, we map Eq.~(\ref{eq:rate_out}) to the M-B model by replacing $\Gamma$ by $\Gamma_{\rm 1D}$ in the second term to project the free-space emission into the 1D mode, and by adding the factor $1/2$ to account for chiral dipole-dipole interactions. This yields the guess 
\begin{equation}
    \gamma_{\rm at}^{\rm st}(\bar{n})\approx \frac{\Gamma_{\rm 1D}}{2}\left(\frac{N^2}{2}\left(\bar{n}-2\bar{n}^2\right)+N \bar{n}\right),
    \label{eq:guess_atomic_emission}
    \end{equation}
where we recognize the coherent and incoherent components by their scaling with $N$. Here, $\bar{n}=\sum_{i=1}^N n_i/N$ is the average excited population in the ensemble. Despite the simplicity of this calculation, Eq.~(\ref{eq:guess_atomic_emission}) is consistent with the M-B results shown in Fig.~\ref{fig:superradiant_emission}a. Of course, the prediction from Eq.~(\ref{eq:guess_atomic_emission}) breaks down at larger optical depths, where the mapping between the Cavity and M-B models is no longer valid. 

\begin{figure}[t]
    \centering
    \includegraphics[width=0.8\linewidth]{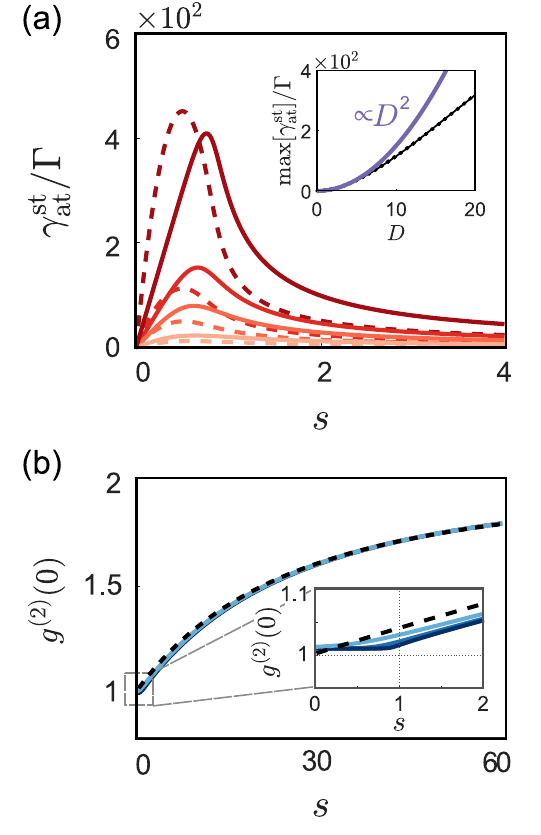}
    \caption{(a) Steady-state atomic emission $\gamma_{\rm at}^{\rm st}$ as a function of the saturation parameter $s$. Solid lines represent the M-B model [Eq.~(\ref{eq:Atomic_emission})] with $D=\{4,8,12,24\}$ (light to dark red). Dashed lines are obtained form the mapping to the Cavity model [Eq.~(\ref{eq:guess_atomic_emission})] for the same optical depths. Inset: Scaling of the steady-state maxima $\max[\gamma^{\rm st}_{\rm at}]$ with $D$. The purple curve is a guide to the eye indicating quadratic scaling $\propto D^2$. (b) Second-order correlations $g^{(2)}(0)$ as a function of $s$ for the optical depths $D=\{4,12,24,40\}$ (light to dark blue). Atomic emission is bunched for $s>1$ and follows a universal curve towards fully chaotic light at very large $s$. The dashed black line is the prediction made from the mapping to the Cavity model [Eq.~(\ref{eq:guess_g2})]. The inset shows the details in the correlations around $s\sim 1$.} 
    \label{fig:superradiant_emission}
\end{figure}
\subsubsection{Second-order photon correlations}
We now discuss the second-order photon correlations in the forward-emitted atomic radiation, which are characterized by
\begin{equation}
    g^{(2)}(0)=\frac{\langle \hat{S}^+ \hat{S}^+\hat{S}^-\hat{S}^-\rangle}{|\langle \hat{S}^+\hat{S}^-\rangle|^2}.
    \label{eq:g2}
\end{equation}
In the Dicke model, and for maximal angular momentum, $g^{(2)}(0)=1$ in the magnetized phase because the steady state is a coherently radiating spin state \cite{Somech2022}. Meanwhile, $g^{(2)}(0)=6/5$ in the unmagnetized phase, which can be calculated by assuming the system is completely mixed within the $j=N/2$ manifold. Similarly, in the M-B model, we expect $g^{(2)}(0)= 1$ in the magnetized regime ($s<1$), where the coherent component dominates the atomic emission. However, we instead expect purely chaotic light $g^{(2)}(0)=2$ deep in the saturated regime ($s\gg1$), since the angular momentum is not conserved and the ensemble becomes a set of independent emitters decaying at random times.

Figure \ref{fig:superradiant_emission}b shows the $g^{(2)}(0)$ from Eq.~(\ref{eq:g2}) calculated using the product state ansatz from Eq.~(\ref{eq:product_state_ansatz}) on Eqs.~(\ref{eq:steady_state_atoms}) and (\ref{eq:steady_state_2}). We observe the predicted behavior above, where correlations go from coherent to chaotic as a function of $s$. Interestingly, the transition between these two regimes sharpens around $s=1$, similar to the magnetization and the transmission. This is consistent with the maxima in Fig.~\ref{fig:superradiant_emission}a, which marks the point where the fraction of incoherent light starts to become relevant. 

We note that the calculated correlations are not exactly $g^{(2)}(0)=1$ in the magnetized regime ($s<1$). This subtlety is due to the incoherent component of the emission. In our mean-field approach, the amount of transmitted incoherent light is over-estimated, because it is assumed that atoms are only driven by the coherent component of the field. Thus, incoherently scattered photons are ``free" to transmit through the system without any additional scattering. We note that in the limit of very weak driving (at most 2 photons sent through the system), $g^{(2)}(0)$ can be calculated by other means, including the effects of re-scattering \cite{Mahmoodian2018}.\\

  \begin{figure*}[t!]
    \centering
    \includegraphics[width=0.97\linewidth]{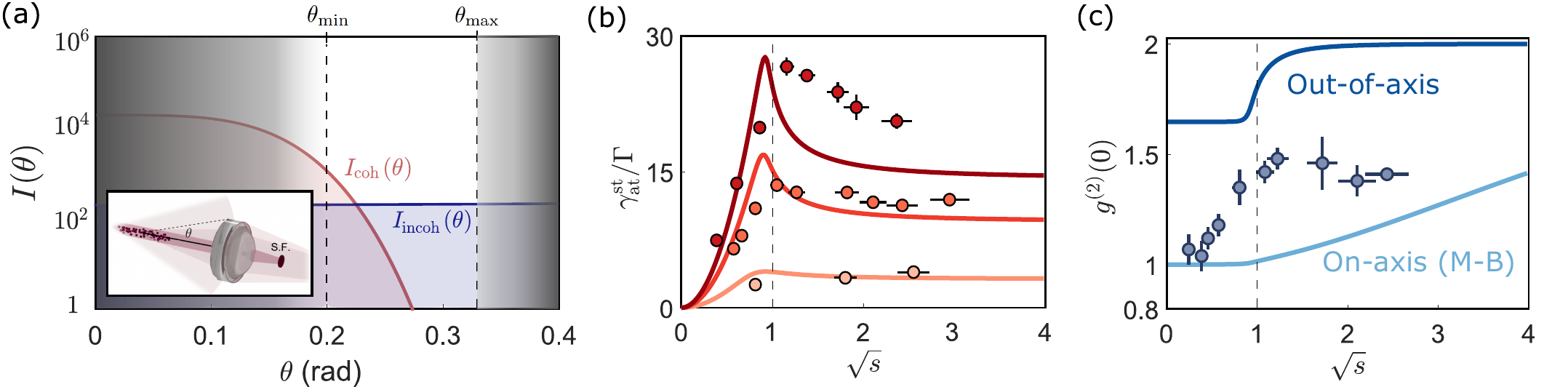}
    \caption{(a) Coherent $I_{\rm coh}(\theta)$ and incoherent $I_{\rm incoh}(\theta)$ intensity components of the atomic emission as a function of the angle $\theta$ between the detected mode $\textbf{k}$ and the optical axis $\textbf{k}_{\rm in}$ defined by the input field. The grey region $\theta\leq\theta_{\rm min}$ represents the range of angles blocked by the spatial filter SF, while the one for $\theta\geq \theta_{\rm max}$ represents the range falling outside the NA of the lens. Here, we take values of $s=1$, $\sigma_L=12.5\lambda$ and $D=20$. (b) Steady-state atomic emission as a function of $\sqrt{s}$ for $D=\{8, 24, 36\}$ (light to dark red). Dashed lines correspond to the modified M-B model and points are the measurements from Ref.~\cite{Ferioli2023}, with the global rescaling discussed in the text. (c) Second-order photon correlations in the atomic emission as a function of $\sqrt{s}$ and for $D=28$. The points are the data from Ref.~\cite{Ferioli2023}, the light blue line is the original (on-axis) M-B model and the dark blue line is the modified prediction when accounting for light measured out of the optical axis.}
    \label{fig:experimental_emission_and_g2}
\end{figure*}

To conclude, we again go back to the Cavity model to derive an approximation for the correlations in the M-B model. We start from the definition of $g^{(2)}(0)$ in Eq.~(\ref{eq:g2}) and substitute the product state ansatz from Eq.~(\ref{eq:product_state_ansatz}). The result can be expressed in terms of sums over the expectation values $\langle \s^z_n\rangle$ and $\langle \s^+_m\rangle \langle\s^+_n\rangle$ with $m\not=n$, which we connect to the coherent $P_{\rm coh}$ and incoherent $P_{\rm incoh}$ components of the atomic emission using Eq.~(\ref{eq:Atomic_emission}). In the limit $N\gg 1$, the $g^{(2)}(0)$ approximates to 
\begin{equation}
   g^{(2)}(0)\approx \frac{2P_{\rm incoh}^2+4P_{\rm incoh}P_{\rm coh}+P_{\rm coh}^2}{(P_{\rm coh}+P_{\rm incoh})^2},
    \label{eq:g2_after_product_state_ansatz}
\end{equation}
to leading order in $N$. From here, we use Eq.~(\ref{eq:guess_atomic_emission}) to replace each component of the atomic emission with their respective guess from the Cavity model, such that $P_{\rm incoh}\approx \Gamma_{\rm 1D}N\bar{n}/2$ and $P_{\rm coh}\approx \Gamma_{\rm 1D}N^2(\bar{n}-2\bar{n}^2)/4$. The result is an expression of $g^{(2)}(0)$ that only depends on the fraction of excited population 
\begin{equation}
  g^{(2)}(0;\bar{n})\approx \frac{\left[N\bar{n}+f(\bar{n})\right]^2-2N^2\bar{n}^2}{\left[f(\bar{n})\right]^2},
    \label{eq:guess_g2}
\end{equation}
where the function $f(\bar{n})=2\gamma_{\rm at}^{\rm st}(\bar{n})/\Gamma_{\rm 1D}$ using Eq.~(\ref{eq:guess_atomic_emission}). It is straightforward to check that Eq.~(\ref{eq:guess_g2}) yields coherent correlations for small non-zero $\bar{n}\ll 1$, and chaotic light for $\bar{n}=1/2$. In practice, the specific  $\bar{n}$ is calculated using the M-B model Eq.~(\ref{eq:steady_state_atoms}) for a given $s$ and $D$.

As we show in Fig.~\ref{fig:superradiant_emission}b, Eq.~(\ref{eq:guess_g2}) reproduces remarkably well the M-B simulations for large $s$. This is because the system becomes spatially homogeneous at large driving intensities ($s\gg1$), and one can represent the ensemble by averaged quantities such as $\bar{n}$. Naturally, this approximation breaks down at lower driving intensities $s\sim 1$, where the ensemble lacks spatial homogeneity and undergoes phase separation, in agreement with the breakdown of Eq.~(\ref{eq:guess_atomic_emission}) observed in Fig.~\ref{fig:superradiant_emission}a for large optical depths.

\subsection{Deviations from the M-B model}
The measurements \cite{Ferioli2023} of the steady-state atomic emission $\gamma_{\rm at}^{\rm st}$ and its photon correlations $g^{(2)}(0)$ deviate significantly from what the M-B model predicts. Indeed, as we will see, the coherent maxima of $\gamma_{\rm at}^{\rm st}$ around $s\sim1$ in Fig.~\ref{fig:superradiant_emission}a are mostly absent, and photon bunching is much stronger than what Fig.~\ref{fig:superradiant_emission}b illustrates. Overall, these discrepancies suggest the coherent component of the atomic emission in the experiment was much weaker than what the M-B model predicts. The fact that the experimental setup fails to detect part of the coherent atomic emission due to filtering~(Fig.~\ref{fig:Setup_measurement_GammaSR}) provides a natural explanation. Here, we present adjusted predictions of the M-B model to qualitatively account for the filtering, such that we can compare with experimental data from Ref.~\cite{Ferioli2023}.

\subsubsection{Atomic emission out of the optical axis}
Our starting point is the experimental setup from Fig.~\ref{fig:Setup_measurement_GammaSR}, where a detector captures the atomic emission $\gamma_{\rm at}$ within a solid angle $\Delta \Theta=2\pi\Delta\theta$. The range of $\theta_{\rm min}<\theta<\theta_{\rm max}$ is limited from below by the spatial filter and from above by the numerical aperture of the lens. For an aspheric lens with NA$=0.5$ \cite{Ferioli2023}, the maximal captured angle is approximately $\theta_{\rm max}\approx 0.35$ rad. Conversely, for a Gaussian beam with $w_0=6.4\lambda$, the angular spread in the paraxial approximation is given by $\theta_{\rm min}\approx \lambda/\pi w_0=0.05$ rad \cite{Ferioli2023}. We note that $\theta_{\rm min}$ was most likely larger than this value, to ensure that no input light reached the detector. While the specific $\theta_{\rm min}$ is unknown, we estimate that it could have been as much as $\theta_{\rm min}\lesssim0.2\text{ rad}$, and assume such a value going forward.

Our goal is to calculate the measured atomic emission by integrating Eqs.~(\ref{eq:Components_intensity_k}), which require specifying $\langle \s^+\s^-\rangle$ and $\langle \s^-\rangle$. To streamline the discussion, we substitute the one-dimensional M-B prediction in Eqs.~(\ref{eq:steady_state_atoms}) for the specific $D$ and $s$, averaged over system length. We note that this is a rough approximation to avoid dealing explicitly with the overly complex diffraction mode in three dimensions. From the discussion surrounding Fig.~\ref{fig:superradiant_emission}a, where the cavity model qualitatively agrees with the original M-B model, we expect that treating the ensemble as a uniform spin wave only introduces quantitative differences for the optical depths we discuss. 

Figure \ref{fig:experimental_emission_and_g2}a illustrates the resulting coherent and incoherent atomic intensities as a function of angle $\theta$, for system parameters $s=1$, $\sigma_L=12.5\lambda$, and $D=20$. We have also shaded the regions $\theta<\theta_{\rm min}$ and $\theta>\theta_{\rm max}$, which are likely excluded from detection. Interestingly, integrating these intensities within the angular spread $\Delta\theta=[0.2, 0.35]$ yields detected coherent and incoherent intensities of the same order already for $s\sim 1$. This is in contrast to the prediction of the M-B model, which assumes that one is detecting on-axis and where the coherent emission should dominate~(see $\theta\sim 0$ region in Fig. \ref{fig:experimental_emission_and_g2}a, where $I_{\rm coh}(\theta)\gg I_{\rm incoh}(\theta)$).
\subsubsection{Steady-state atomic emission $\gamma_{\rm at}^{\rm st}$}
We use the previous assumptions to discuss the measurements of the steady-state atomic emission $\gamma_{\rm at}^{\rm st}$ from Ref.~\cite{Ferioli2023}. The data are shown in Fig.~\ref{fig:experimental_emission_and_g2}b as a function of $s$ and varying $D$. Compared to what the M-B model predicts in Fig.~\ref{fig:superradiant_emission}a, the large coherent maxima are mostly absent. Moreover, for drivings $s>1$, the atomic emission appears to stabilize, suggesting saturation is achieved much faster than in Fig.~\ref{fig:superradiant_emission}a. 

Interestingly, these features can be qualitatively reproduced with our modified M-B model. Following the procedure used in Fig.~\ref{fig:experimental_emission_and_g2}a, we use Eq.~(\ref{eq:Components_intensity_k}) to calculate the coherent and incoherent atomic intensities. Then, we calculate the out-of-axis atomic emission by integrating the resulting intensities within the interval of captured angles $\Delta\theta\sim[0.2, 0.35]$ rad. The result is shown in Fig.~\ref{fig:experimental_emission_and_g2}b with solid lines, where this simple model accounting for the different transverse profiles of coherent/incoherent emission seems sufficient to produce consistency with the experimental results. According to this interpretation, the incoherent component dominates the atomic emission already for $s\sim 1$.

We note that measuring an absolute value for $\gamma_{\rm at}$ is challenging, as it requires precisely estimating the coupling efficiency to the fiber detection mode, the filtering of the input field, or the mechanical displacement of the cloud at large intensities, among other effects  \cite{Ferioli2023}. For that reason, the experimental values for $\gamma_{\rm at}$ in Figs.~\ref{fig:experimental_emission_and_g2}b have been effectively rescaled by an arbitrary global factor. 

Due to the number of experimental uncertainties and the simplified manner in which we try to incorporate three-dimensional effects into the 1D M-B model, we emphasize that our model should not be taken as a full, quantitatively accurate description of the three-dimensional propagation effects. However, we believe that it should qualitatively capture the key physics and identifies an important subtlety of measuring the forward atomic emission.

\subsubsection{Second-order photon correlations $g^{(2)}(0)$}
Similarly, Fig.~\ref{fig:experimental_emission_and_g2}c shows the measurements of the $g^{(2)}(0)$~(circles) from Ref.~\cite{Ferioli2023}. In the plot, we also include the original M-B model prediction~(light blue curve). The correlations measured in the magnetized regime ($s<1$) are slightly bunched, while $g^{(2)}(0)\approx1$ in the M-B model. Moreover, in the unmagnetized regime ($s>1$), the bunching in the experimental data is much higher than what the M-B model predicts. 

Again, these discrepancies can possibly be explained by assuming the detection scheme. Starting from the mean-field Eq.~(\ref{eq:g2_after_product_state_ansatz}), we substitute the coherent and incoherent components by the integrals of Eqs.~(\ref{eq:Components_intensity_k}) over the same range of angles $\Delta\theta=[0.2, 0.35]$. Plotting this result in Figure \ref{fig:experimental_emission_and_g2}c~(dark blue curve), we see that Eq.~(\ref{eq:g2_after_product_state_ansatz}) predicts an increase in the bunching, which stems from the additional purely incoherent light having chaotic correlations $g^{(2)}(0)=2$. The details of the blue curve are highly sensitive to choices of parameters in our model, but nonetheless this curve provides a plausible picture of the origin of the correlations seen in the experiment.

\section{Conclusion}
Here, we have investigated the properties of a driven, elongated atomic ensemble using a model based on the Maxwell-Bloch (M-B) equations. At the mean-field level, we identified non-analytic behavior in observables in the thermodynamic limit ($D\rightarrow \infty$), such as the steady-state transmission and magnetization, resembling that of the driven-dissipative Dicke phase transition. This non-analyticity in free space, however, stems from a lack of spatial homogeneity, which in turn results from propagation effects absent in a cavity setup. Physically, the ensemble splits into saturated and unsaturated regions, a phenomenon closer to what we term ``phase separation" rather than a conventional phase transition.  

Under specific conditions, such as low optical depth or large driving strength, the propagation effects become negligible and the mean-field equations of the M-B model become equivalent to those of the Cavity model. This mapping provides a microscopic justification for the similarities between the measurements in Ref.~\cite{Ferioli2023} and the driven Dicke model. Nevertheless, we have shown that the M-B model accurately reproduces most of the experimental data.

Moving forward, our work holds direct relevance for state-of-the-art experimental setups \cite{Ferioli2023, Lechner2023}. Inspired by the mapping between M-B and Cavity models, it would be interesting to explore the conditions under which cavity phenomena, such as superradiant lasing~\cite{Jager2021,Shankar2021,Ferioli2021_PRL_superradiance}, spin squeezing~\cite{Puri1979,Tudela2013} or time crystalline behaviour~\cite{Iemini2018}, manifest in systems with propagation effects like free space atomic ensembles or chiral waveguide QED setups. Specifically, it would be useful to quantify the efficiency of such applications -- e.g. scaling with optical depth -- before the phase separation and other effects limit their performance. Similarly, it would also be interesting to extend the discussion of this mapping beyond mean-field arguments \cite{Kasper2023,Robicheaux2021}.

During the completion of our work, we became aware of a related recent work that reaches similar conclusions as us using a complementary theoretical treatment \cite{Argawal_2024}.

\section{Acknowledgements}
The authors acknowledge stimulating discussions with Fernando de Iemini, Jamir Marino, Francis Robicheaux, and Ana Maria Rey. D.G. acknowledges support from the Secretaria d’Universitats i Recerca de la Generalitat de Catalunya and the European Social Fund (2020 FI B 00196). L.B. acknowledges support from the European Union’s Horizon Europe research and innovation program under Grant Agreement No. 101113690 (PASQuanS2.1). E.S. acknowledges support from the Israel Science Foundation (ISF), the Center for New Scientists at the Weizmann Institute of Science, and the Council for Higher Education (Israel). D.E.C. acknowledges support from the European Union, under European
Research Council grant agreement No 101002107 (NEWSPIN), FET-Open grant agreement
No 899275 (DAALI) and EIC Pathfinder Grant No 101115420 (PANDA); the Government of
Spain under Severo Ochoa Grant CEX2019-000910-S (MCIN/AEI/10.13039/501100011033);
Generalitat de Catalunya (CERCA program and AGAUR Project No. 2021 SGR 01442);
Fundació Cellex, and Fundació Mir-Puig.
This work is also supported by
the Agence Nationale de la Recherche 
(ANR-22-PETQ-0004 France 2030, project QuBitAF), 
and the European Research Council (Advanced grant No. 101018511-ATARAXIA)
and the Horizon Europe programme HORIZON-CL4-2022-QUANTUM-02-SGA via the project 101113690 (PASQuanS2.1). 

\bibliography{Biblio}

\appendix

\section{Time dynamics of atomic emission} \label{sec:Initial_peak}
Here, we complement the discussion in Sec. IVB in the main text on the atomic emission $\gamma_{\rm at}$ by discussing its dynamics at early times. Specifically, we numerically solve the time evolution of the discrete Eqs.~(\ref{eq:Maxwell_Bloch_continuum}) within mean-field approximation and starting with all atoms in their ground state. Then, we calculate the atomic emission at different times using Eq.~(\ref{eq:Atomic_emission}). The results below are obtained with just the original M-B model, i.e. without any additional correction for the measurement being off-axis (like in Sec.~IV-C). 

\begin{figure}[t!]
    \centering
    \includegraphics[width=0.95\linewidth]{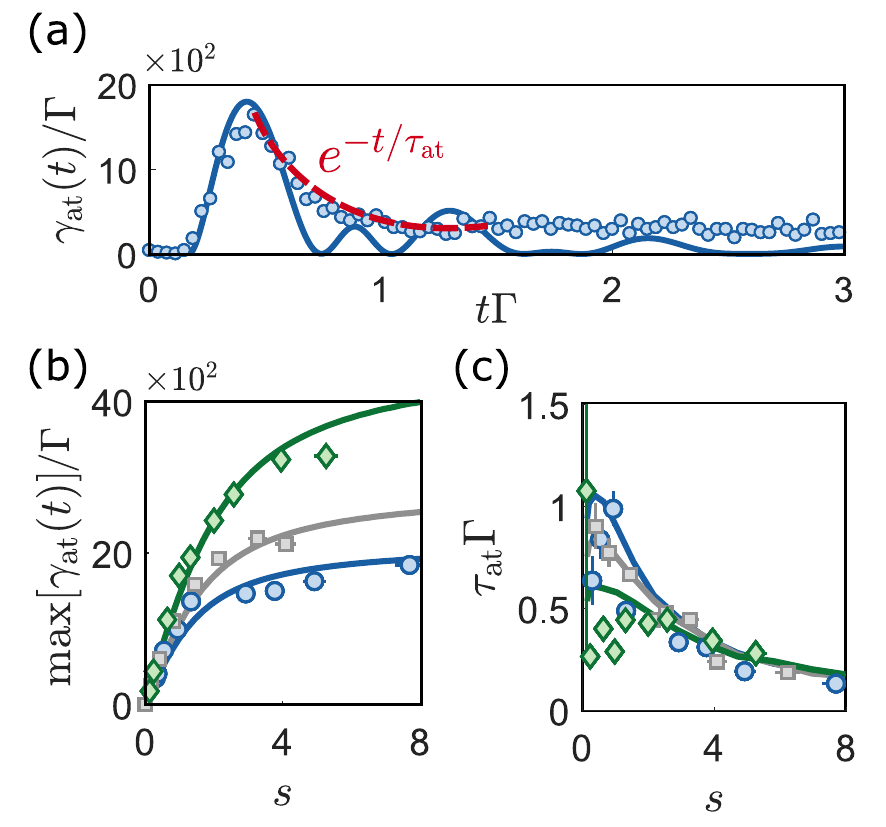}
    \vspace{-1em}
    \caption{(a) Atomic emission $\gamma_{\rm at}(t)$ exhibits a large coherent peak at early times that rapidly relaxes into a plateau. The solid line is a numerical simulation of the M-B model, while the points are experimental measurements using the setup from Ref.~\cite{Ferioli2023}. We have rescaled the experimental data as indicated in the main text. Here, $s=5$ and $D=26$. (b,c) Peak height $\max[\gamma_{\rm at}(t)]$ and damping rate $\tau_{\rm at}$ as a function of the saturation parameter $s$ for $D=[26, 30, 38]$ (blue, gray and green, respectively). We again use solid lines for the M-B model and points for the experimental data. The data in (b) has been rescaled as indicated in the main text.}
    \label{fig:Initial_peak}
\end{figure}

In Fig.~\ref{fig:Initial_peak}a, we plot $\gamma_{\rm at}(t)$ as a function of time for a particular realization ($s=5$, $D=26$). At very early times  ($t\Gamma< 1$), the atomic emission exhibits a prominent coherent peak, which is rapidly damped into a steady-state plateau. Since the time duration of the peak is short, dipole-dipole interactions cannot significantly influence the dynamics. Therefore, this coherent peak emerges solely from a combination of Rabi oscillations and constructive interference. 

Figure~\ref{fig:Initial_peak}a also shows the experimental measurements of $\gamma_{\rm at}(t)$ for the same realization, obtained with the setup in Fig.~\ref{fig:Setup_measurement_GammaSR}. Here, the experimental values have been effectively rescaled by an arbitrary global factor as in Fig.~\ref{fig:experimental_emission_and_g2} in the main text for the same reasons. In other words, Fig.~\ref{fig:Initial_peak}a shows a good qualitative agreement between the experiment and the M-B model, despite the possibility of their peak heights being different. The discrepancies at later times are probably related to the atomic emission being measured out of the optical axis, and they could be corrected by modifying the M-B model following the discussion in Sec.~IV-C.

We now explore the properties of the coherent maxima as a function of optical depth $D$ and saturation parameter $s$. In Figs.~\ref{fig:Initial_peak}b,c, we plot the peak height $\gamma_{\rm max}=\max[\gamma_{\rm at}(t)]$ and damping rate $\tau_{\rm at}$ as a function of the saturation parameter $s$. The damping rate is obtained by fitting the function $\gamma_{\rm at}(t)=\gamma_{\rm max}\exp(-t/\tau_{\rm at})$ to the interval right after the intensity maximum. Overall, their scalings with the driving intensity are well captured by the M-B model, despite not including any additional correction for the light being measured off-axis.

\section{Dipole-dipole interaction matrix elements} \label{sec:Hddelements}
Here, we present a complementary calculation for the mapping between the M-B and Cavity models at large driving strengths. Specifically, we study analytically how the coherent dipole-dipole interaction $\hat{H}_{\rm dd}$ in Eq.~(\ref{eq:H_dd}) couples different angular momentum states. To do so, we calculate how $\hat{H}_{\rm dd}$ acts on the total angular momentum basis $|j,m\rangle$. We note this is not trivial since $\hat{H}_{\rm dd}$ contains only local operators that break the degeneracy of $|j,m\rangle$.

Since each term in $\hat{H}_{\rm dd}$ acts on a pair of atoms at a time, it is convenient to split the atomic ensemble into two subsystems: an arbitrary two-atom pair $(n,l)$ and the remaining $N-2$ atoms. We then expand a generic $N$ atom state $|j,m\rangle$ into the basis constructed from the tensor product between the basis of each subsystem, i.e. $|j_{nl}, m_{nl}\rangle \otimes |j_{N-2}, m_{N-2}\rangle$, such that
\begin{multline}
        |j,m\rangle  = \sum_{\{j_{N-2}, m_{N-2}\} \in \mathcal{C}} |0,0; j, m\rangle\  +\\ +  C^{1,0, j_{N-2}, m}_{J,m}|1,0; j_{N-2}, m_{N-2}\rangle  + \\
     + C^{1,1, j_{N-2},m_{N-2}}_{j,m}|1,1; j_{N-2},m_{N-2}\rangle  +  \\
    +C^{1,-1, j_{N-2},m_{N-2}}_{j,m}|1,-1; j_{N-2},m_{N-2}\rangle , 
     \label{eq:JJzstate}
\end{multline}
where the sum is restricted over the $j_{N-2}$ and $ m_{N-2}$ that satisfy the angular-momentum composition rules $\mathcal{C}$. In Eq.~(\ref{eq:JJzstate}), we have defined the Clebsch-Gordan coefficient $C^{j_{nl},m_{nl}, j_{N-2},m_{N-2}}_{j,m} \equiv  \langle j_{nl}, m_{nl}; j_{N-2}, m_{N-2}|j, m\rangle$, and used the relations 
\begin{subequations}
    \begin{equation}
        C^{0,0, j', m'}_{j,m} = \delta_{j, j'} \delta_{m, m'},
    \end{equation}
    \begin{equation}
        C^{1,0,j', m'}_{j,m} = C^{1,0, j', m'}_{j,m} \delta_{m, m'},
    \end{equation}
\end{subequations}
Here, we have simplified the notation by removing the label of the atom pair $(n,l)$.\\

Having defined the Clebsch-Gordan coefficients between the basis, we now calculate how the element ${\hat{H}_{{\rm dd}, nl} \equiv \sigma^+_n \sigma^-_l - \sigma^+_l \sigma^-_n}$, appearing in the coherent dipole-dipole interaction ${\hat{H}_{\rm dd}= -i\hbar\Gamma_{\rm 1D}/4 \sum_{n>l}\hat{H}_{\rm dd, nl}}$, acts on the angular momentum states. For simplicity, we focus only on the maximal angular momentum states $j=N/2$, obtaining
\begin{equation}
    \hat{H}_{\rm dd} | N/2, m\rangle  = i \hbar\Gamma_{\rm 1D} F(m) |\psi\rangle.
    \label{eq:Hdd_onJJz_state}
\end{equation}
Here, we have defined
\begin{equation}
    F(m) |\psi\rangle = \frac{C^{1,0, \left(N/2-1\right),m}_{N/2,m}}{4} \sum_{l > m} |0,0;(N/2-1),m\rangle_{n,l},
\end{equation}
where we exploited the permutational symmetry of the $j=N/2$ states and imposed $\langle \psi|\psi\rangle = 1$. 

\begin{figure}
    \centering
    \includegraphics[width=0.9\linewidth]{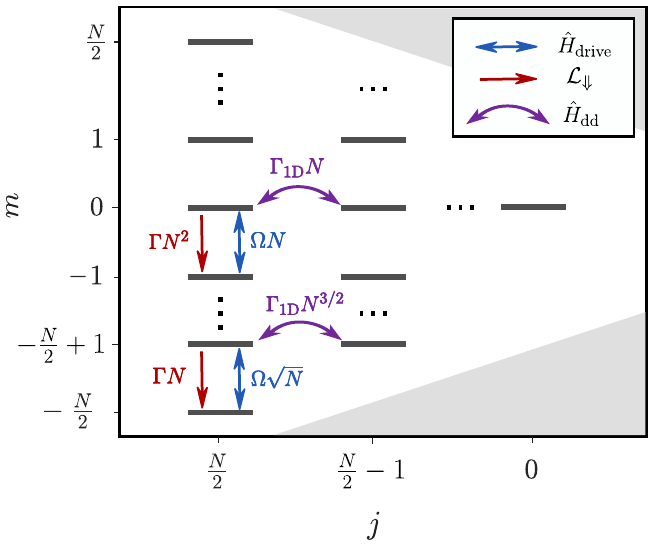}
    \caption{Matrix elements of the coherent driving Hamiltonian $\hat{H}_{\rm drive}$ (blue arrows), of the collective dissipation $ \mathcal{L}_{\Downarrow}$ (red arrows) and of the coherent dipole-dipole interactions $\hat{H}_{\rm dd}$ (violet arrows) for Dicke states with $m=0$ and $m=-N/2+1$.}
    \label{fig:Hddmatrix}
\end{figure}

Using the Clebsh-Gordon coefficients defined above, one can show that $\hat{H}_{\rm dd}$ couples the state $|N/2,m\rangle$ with the corresponding unique state $|\psi\rangle$ in the $j=N/2-1$ manifold with strength
\begin{equation}
     F(m) = \sqrt{\frac{(N/2+m) \left(N/2-m\right) \left(N+1\right)}{24}}.
     \label{eq:Appendix_scalings_Lisa}
\end{equation}
Equation (\ref{eq:Appendix_scalings_Lisa}) provides the scaling of the coherent dipole-dipole matrix elements with system size. In the thermodynamic limit $N\rightarrow\infty$, the coupling scales linearly with atom number $F\propto N$ at the extremes of the $j=N/2$ manifold ($|m|\approx N/2$), while $F \propto N^{3/2}$ at the center ($m\approx 0$). Next, we identify two distinct regimes depending on how the driving strength and collective decay compare to these scalings, as we show in Fig.~\ref{fig:Hddmatrix}. 

For weak drivings ($ \Omega \lesssim \Gamma\sqrt{N}$), the effect of coherent dipole-dipole interactions is not negligible, as $\hat{H}_{\rm dd}$ mixes states from different angular momentum manifolds even for small number of excitations. This suggests that propagation effects matter, and is consistent with both our purity discussion based on the lateral decay rates (Sec.~\ref{Sec:lateral_decay_rates}) and the discrepancy between the Cavity and M-B models in Sec.~\ref{Sec:Mapping_cavity_and_MB}. 

On the other hand, for strong drivings ($\Omega \sim \Gamma N$), the effect of $\hat{H}_{\rm dd}$ becomes negligible for experimentally realistic cooperativities ($\Gamma\gg\Gamma_{\rm 1D}$). This result complements the intuition developed in Sec.~\ref{Sec:Mapping_cavity_and_MB}, where the Cavity and M-B models converge as propagation effects vanish.

\end{document}